\documentclass[a4paper,11pt]{article}
\pdfoutput=1
\usepackage{jheppub}
\usepackage[T1]{fontenc}
\usepackage{graphicx}
\usepackage{amsmath}
\usepackage{amsmath}
\usepackage{xspace}
\usepackage{subfig}
\usepackage{color}
\usepackage[utf8]{inputenc}
\usepackage{commath}
\usepackage[utf8]{inputenc} 

\newcommand{\mzprime}{$m_{Z^{\prime}}$\,}
\newcommand{\zprime}{$Z^{\prime}$\,}
\newcommand{\gprime}{$g_{B-L}$\,}

\title{\boldmath Searching for a light $Z'$ through\\Higgs production at the LHC}

\author[a, b]{Frank F. Deppisch,}
\author[b]{Suchita Kulkarni,}
\author[a]{Wei Liu}

\affiliation[a]{University College London, Gower Street, London WC1E 6BT, UK}
\affiliation[b]{Institut f{\"u}r Hochenergiephysik, {\"O}sterreichische Akademie der Wissenschaften,\\ Nikolsdorfer Gasse 18, 1050 Wien, Austria}

\emailAdd{f.deppisch@ucl.ac.uk}
\emailAdd{suchita.kulkarni@oeaw.ac.at}
\emailAdd{wei.liu.16@ucl.ac.uk}

\abstract{We investigate the potential of LHC resonance searches in leptonic final states to probe the $Z'$ in the minimal $U(1)_{B-L}$ model. Considering the current constraints on the \zprime in terms of its mass $m_{Z'}$ and the associated gauge  coupling \gprime as well as constraints in the Higgs sector, we analyse the potential of dilepton and four lepton final states for \zprime production. This includes Drell-Yan production, Higgs mediated decays and final state radiation processes concentrating only on the ATLAS and CMS detectors at the LHC. We show that the four-lepton final state is sensitive to $m_{Z'}$ as low as 0.25~GeV. Furthermore, setting the Higgs mixing to $\sin\alpha = 0.3$, this final state has a strong sensitivity and it probes regions of parameter space where the \zprime is long-lived. We demonstrate the sensitivity at the High Luminosity LHC and comment on the potential of probing displaced vertices due to long-lived \zprime. Finally, we also comment on the strength of \zprime and Higgs mediated heavy neutrino processes by taking into account the constraints derived.} 

\notoc

\begin{document} 
\maketitle
\flushbottom

\section{Introduction}
\label{sec:intro}
The presence of finite yet small tiny masses of neutrinos remains as one of the puzzles within the Standard Model (SM). Specifically, the ultimate goal is to determine the nature of neutrinos and the corresponding mechanism of neutrino mass generation. The mechanism is typically believed to be accompanied by the breaking of lepton number $L$ symmetry resulting in a Majorana neutrino character. At the LHC and in other searches, it can be probed by searching for heavy neutrinos (or neutral heavy leptons, to use the other often used name) and other mediators of the different types of the seesaw mechanism. The difficulty to probe a parameter space relevant for light neutrino mass generation is often challenging due to the required lightness of neutrino; generically, this either demands heavy mediators (which may not be accessible at colliders) or small couplings to the SM neutrinos (which suppresses the mediator production rates). Other solutions exist, though, such as in inverse seesaw scenarios \cite{Mohapatra:1986bd}, where the suppression is achieved through an weakly broken lepton number symmetry, or in radiative models with loop-suppressed neutrino masses.

Moreover, in the prominent seesaw type-I mechanism where three right-handed neutrinos $N_i$ are added to the SM, the $L$ symmetry breaking is explicit by assuming Majorana masses for the right-handed neutrinos. While this is perfectly valid as such masses for the gauge-sterile right-handed neutrinos are not forbidden by the SM gauge symmetry, the question of where the light neutrinos get their mass is simply shifted: where do the right-handed neutrinos get their masses? Clearly, the observable presence of the heavy sterile neutrinos provides the crucially testable consequence but embedding the seesaw mechanism in a more complete model will also provide additional means to probe the mechanism of neutrino mass generation. Arguably one of the simplest ultraviolet (UV) complete models for this purpose is described by the $U(1)_{B-L}$ extension of the SM gauge group \cite{Davidson:1978pm, Mohapatra:1980qe}, where particles are additionally charged under the quantum number $B-L$ ($B$ is the usual SM baryon number). Here, three right-handed Majorana neutrinos are added as well, in order to give masses to the light neutrinos via seesaw type-I but also to cancel anomalies. The right-handed neutrino Majorana masses are generated by the spontaneous breaking of the $U(1)_{B-L}$ symmetry via an extra Higgs field $\chi$. 

The important prediction of this model is the presence of an additional gauge boson \zprime associated with the $B-L$ gauge symmetry. The $Z'$ can be probed for in several different ways. LHC searches for heavy resonance in dilepton final states put a strict bound on $m_{Z'} > 4.5$~TeV~\cite{Aaboud:2017buh} for a $g_{B-L}$ coupling similar to that of the SM $Z$ boson. While the $B-L$ breaking scale $\langle\chi\rangle = m_{Z^\prime}/(2g_{B-L})$ is constrained to be larger than 3.45~TeV from LEP-II~\cite{LEP:2003aa, Anthony:2003ub, Carena:2004xs, Cacciapaglia:2006pk}, these limits are not applicable when \mzprime becomes too small. Neutrino scattering experiments set an effective limit on $\langle\chi\rangle \gtrsim 1$~TeV \cite{Harnik:2012ni,Bellini:2011rx,Bauer:2018onh, Vilain:1993kd, Deniz:2009mu}. The wider $m_{Z'} - g_{B-L}$ parameter space can for example be explored using the Constraints On New Theories Using Rivet ({\tt CONTUR}) method for $m_{Z'} \gtrsim 1$~GeV incorporating ATLAS and CMS results \cite{Amrith:2018yfb, Butterworth:2016sqg}. For $m_{Z'} < 10$~GeV limits are set at $g_{B-L} \lesssim 10^{-4}$ to $10^{-3}$ from recasting dark photon searches at LHCb using {\tt Darkcast}~\cite{Ilten:2018crw}. For even smaller $m_{Z^\prime} < 1$~GeV, proton and electron beam dump experiments are sensitive to long-lived $Z^\prime$ for sufficiently small $g_{B-L} \sim 10^{-8}-10^{-4}$ \cite{Ilten:2018crw}, cf. our summary Fig.~\ref{fig:darkcast_summary}.

Despite intense efforts to constrain new resonances at colliders, within the $B-L$ model, the parameter space of \zprime masses between 1 to 100~GeV remains relatively unconstrained for \gprime  $< 10^{-3}$~\cite{Ilten:2018crw}. In this work, we concentrate on this parameter space, and analyse the reach of existing searches in leptonic final states at the ATLAS and CMS detectors. For masses less than 10 GeV, the factorization theorem is no longer applicable and the production should be dealt with via e.g. the Vector Dominance Mechanism. An alternative way to look for low mass \zprime is to explore their production via heavier resonances. For $B-L$ model this could be production via the $B-L$ Higgs or the SM Higgs. As we will demonstrate later, the $B-L$ Higgs is not a good production channel however \zprime production via the SM Higgs remains a viable option. Recently this production mechanism is under attention as one of the key process to explore dark photon models at the LHC. Due to new developments in analysis strategies, dark photons masses as low as 0.25 GeV are constrained. This motivates analysis of $B-L$ models in the same final state and understanding the reach of these searches. Above the mass of 10 GeV,  the resonance searches in e.g. dilepton final states will prove to be useful. Currently, the best limits in this region are obtained via the LHCb search for dark photons~\cite{Aaij:2017rft}.

\zprime production via SM Higgs as will be explored in this work is however dependent on the mixing angle between the $B-L$ Higgs and the SM Higgs. This production mechanism is therefore subject to constraints on the Higgs sector from both direct and indirect searches~\cite{Bechtle:2013xfa, Ilnicka:2018def, Bechtle:2014ewa, ATLAS:2016gld, ATLAS:2016oum, ATLAS:2016pkl, ATLAS:2016awy, ATLAS:2016nke, ATLAS:2016ldo, Sirunyan:2017exp, CMS:2016ixj, CMS:aya, CMS:bxa, Khachatryan:2015cwa, CMS:2017vpy, Aaboud:2017rel, Lopez-Val:2014jva, Robens:2015gla}. These limit the $B-L$ Higgs -- SM Higgs mixing angle and therefore the strength of the \zprime production at the LHC. Current constraints on the Higgs mixing angle include those from direct searches for additional Higgs bosons at the LHC, the SM Higgs signal strength measurements, as well as constraints from electroweak observables and the measurements of the $W$ mass. They are further complemented by constraints from theoretical considerations of the perturbativity of the Higgs couplings, unitarity and vacuum stability.

In this work, we explore three different \zprime production mechanisms and derive limits for low mass \zprime. The first process we consider is $pp \to Z'\to \mu^+\mu^-$. We refer to this as the Drell-Yan \zprime production channel. Second, we consider the final state radiation of \zprime. Here the \zprime is radiated off via the muons in decay products of SM Z. More precisely, the process is $pp \to Z \to \mu^+\mu^- Z' \to 4\mu$. Finally, we consider the \zprime production via SM Higgs portal with \zprime decays to leptonic final state $pp \to h \to Z' Z' \to 4 l$. In combining these three processes, we derive new constraints in the $m_{Z'}$ -- \gprime parameter space of the minimal $U(1)_{B-L}$ model. 

As it will turn out, the $Z'$ will be long-lived for masses $m_{Z'} \lesssim 1$~GeV and $g_{B-L} \lesssim 10^{-5}$. At the LHC, this lifetime frontier can be explored via displaced signatures. This region is of particular interest for Higgs mediated \zprime production. Due to the large mass difference between the SM Higgs and the \zprime, the \zprime receives a large boost leading to macroscopic lab frame displacement and displaced vertices can be observed. We will carefully chalk out the regions where \zprime is displaced and demonstrate the potential of existing searches.

The plan of the paper is as follows: In section~\ref{blreview}, we briefly review the minimal $B-L$ model and its parameter space under consideration. Section~\ref{param_space} contains a discussion of the associated collider signatures, whereas Section~\ref{recast} is devoted to a discussion of the LHC searches we incorporate in our analysis. We derive limits in Section~\ref{results} and Section~\ref{conclusion} concludes our work.

\section{The Minimal $U(1)_{B-L}$ Model}
\label{blreview}

\subsection{Model Setup}

The minimal $U(1)_{B-L}$ was first described in Ref.~\cite{Mohapatra:1980qe}. We here discuss the salient features as far as relevant for our discussion. In addition to the particle content of the SM, the $U(1)_{B-L}$ model incorporates an Abelian gauge field $B^\prime_\mu$, a SM singlet scalar field $\chi$ and three RH neutrinos $N_i$. The gauge group is $SU(3)_C\times SU(2)_L \times U(1)_Y \times U(1)_{B-L}$, where $\chi$ and $N_i$ have $B-L$ charges $B-L = +2$ and $-1$, respectively. The SM fermions have $B-L$ charges determined by their usual baryon $B$ and lepton $L$ numbers whereas all other SM fields are uncharged under $U(1)_{B-L}$. This fully describes the gauge sector of the model, where we make the assumption that the mixing between the $U(1)_{B-L}$ and $U(1)_Y$ fields vanishes. Even though this kinetic mixing arises naturally in loop diagrams, as it is scale dependent, we assume it to be zero at the electroweak scale, and the value at other scales can be derived from the renormalisation group evolution, cf. Ref.~\cite{DEramo:2017zqw} for a similar example. This assumption is made as a simplification to analyze the interplay between the Higgs mixing and the $U(1)_{B-L}$ gauge coupling.

The scalar sector is uniquely determined by the scalar potential
\begin{align}
\label{VHX}
	{\cal V}(H,\chi)
	= m^2 H^\dagger H + \mu^2 |\chi|^2 + \lambda_1 (H^\dagger H)^2 
	+ \lambda_2 |\chi|^4 + \lambda_3 H^\dagger H |\chi|^2,
\end{align}
incorporating all allowed terms for the SM Higgs doublet $H$ and the new scalar field $\chi$. The breaking of the $(B-L)$ symmetry is achieved spontaneously such that $\chi$ acquires a vacuum expectation value (VEV) $\langle\chi\rangle$ breaking  $SU(3)_C\times SU(2)_L \times U(1)_Y \times U(1)_{B-L} \to SU(3)_C\times SU(2)_L \times U(1)_Y$ above the electroweak (EW) scale. Consequently, the $U(1)_{B-L}$ gauge field acquires a mass
\begin{align}
\label{eq:zprimemass}
	m_{Z'} = 2 g_{B-L} \langle\chi\rangle.
\end{align} 

Likewise, the $U(1)_{B-L}$ and EW breaking will generate a mixing between $\chi$ and the SM Higgs through the $\lambda_3$ term in Eq.~\eqref{VHX}. Specifically, the mass matrix of the Higgs fields $(H, \chi)$ at tree level is \cite{Robens:2015gla}
\begin{align}
\label{mass}
	M_h^2 = \begin{pmatrix}
		2\lambda_1 v^2 & \lambda_3 x v \\
		 \lambda_3 x v & 2\lambda_2 x^2
	\end{pmatrix},
\end{align}
with $x = \langle\chi\rangle$ and $v = \langle H_0 \rangle$, resulting in the mass eigenstates $h$, $h_\chi$ with masses 
\begin{align}
\label{Higgsmass}
	m^2_{h (h_\chi)} 
	= \lambda_1 v^2 + \lambda_2 x^2 
	  -(+) \sqrt{(\lambda_1 v^2 - \lambda_2 x^2)^2 + (\lambda_3 xv)^2}.
\end{align} 
Here we assume that the SM-like Higgs $h$ is lighter than the exotic Higgs $h_\chi \sim \chi$, and the physical Higgs states ($h, h_\chi)$ are related to the gauge states ($H, \chi$) as
\begin{align}
\label{Higgs mixing}
	\begin{pmatrix}
		h \\ h_\chi
	\end{pmatrix} = 
	\begin{pmatrix}
		\cos\alpha & -\sin\alpha \\
		\sin\alpha &  \cos\alpha
	\end{pmatrix}
	\begin{pmatrix}
		H \\ \chi
	\end{pmatrix}.
\end{align} 
Here, $\alpha$ is the mixing angle relating the two bases. At tree-level it can be computed from the parameters in the scalar potential and the scalar VEVs as
\begin{align}
\label{lambda}
	\tan(2\alpha) = \frac{\lambda_3 v x}{\lambda_2 x^2 - \lambda_1 v^2}.
\end{align}

While we do not discuss heavy neutrinos explicitly in this paper, we include for completeness how the seesaw type-I mechanism is naturally embedded in this model. Because of the charge assignments $(B-L) = -1$ for the right-handed neutrinos $N_i$ and $(B-L) = +2$ for the scalar $\chi$, the following two Yukawa-type interactions are allowed by the model gauge group,
\begin{align}
\label{LY}
	{\cal L} \supset 
	   - y_{ij}^\nu \overline{L_i}\nu_{Rj}\tilde{H}
	   - y_{ij}^M \overline{\nu^c_{Ri}} \nu_{Rj}\chi 
	   + \text{h.c.}.
\end{align} 
Here, $L_i$ are the SM lepton doublets, $\tilde{H} = i\sigma^2 H^\ast$ and a summation over the generation indices $i, j = 1, 2, 3$ is implied. The Yukawa matrices $y^\nu$ and $y^M$ are a priori arbitrary; the RH neutrino mass is generated due to breaking of the $B-L$ symmetry, with the mass matrix given by $M_R = y^M \langle\chi\rangle$. The light neutrinos mix with the RH neutrinos via the Dirac mass matrix $m_D = y^\nu v/\sqrt{2}$. The complete mass matrix in the $(\nu_L, \nu_R)$ basis is then
\begin{align}
\label{MD}
	{\cal M} = 
	\begin{pmatrix}
		0   & m_D \\
		m_D & M_R
\end{pmatrix},
\end{align} 
In the seesaw limit, $M_R \gg m_D$, the light and heavy neutrino masses are $m_\nu = - m_D M^{-1}_R m^T_D$ and $ m_N = M_R$, respectively. Considering a simple one-generational scenario, this leads to the celebrated seesaw mechanism which induces a mixing between the light and heavy neutrinos, 
\begin{align}
\label{Rotation}
	\begin{pmatrix}
		\nu_L \\ \nu^c_R
	\end{pmatrix} \approx
	\begin{pmatrix}
    	1 & -\theta_\nu \\
		\theta_\nu &  1
	\end{pmatrix}
	\begin{pmatrix}
		\nu \\ N
	\end{pmatrix},
\end{align}
with the small mixing angle $\theta_\nu = \sqrt{m_\nu / m_N}$. For $m_\nu \approx 0.1$~eV and $m_N \approx 10$~GeV this gives a very small mixing angle $\theta_\nu \approx 3\times 10^{-6}$. With these choices of parameters, one can compute the Yukawa couplings for the light and heavy neutrinos, $y^M = m_N/\langle \chi \rangle \approx 10^{-3}$, and $y^\nu \approx \sqrt{m_N m_\nu}/v \approx 10^{-7}$. Considering only SM gauge interactions, heavy neutrino production rates via the SM $W$ and $Z$ will be suppressed by $\theta_\nu^2$. In the $(B-L)$ model considered here, the heavy neutrinos are also produced via $Z'$ and $\chi$. It is therefore important to probe these exotic particles as stringently as possible to learn about the viability of neutrino mass generation mechanisms near the EW scale.

\subsection{Constraints on the Parameter Space}
The main focus of this work is to explore the viability of producing $Z'$ through the SM Higgs. In this context, we are specifically interested in three model parameters: the $Z'$ mass $m_{Z'}$, the $U(1)_{B-L}$ gauge coupling $g_{B-L}$ and the Higgs mixing angle parametrized as $\sin\alpha$. As already stated, we assume that the mixing of the $Z'$ with the SM $Z$ vanishes and we implicitly consider the second Higgs $h_\chi$ to be heavy enough so as not to affect our calculations. 

Neutral gauge bosons, such as the $Z'$ in our model, have been searched for in numerous experiments. As the \zprime couples to quarks and leptons at tree level, it can be searched for via the s-channel production at various colliders. Several such searches exist at e.g. KLOE~\cite{Anastasi:2015qla}, BaBar~\cite{Lees:2014xha} and at the LHC~\cite{Aaij:2017rft}. Resonance searches at the LHC for additional gauge bosons in dilepton final states currently rule out \zprime masses up to approximately 4 TeV~\cite{Sirunyan:2018exx, Aad:2019fac}. These searches however are limited in their ability to search for \zprime below a 100 GeV due to large event rates at the LHC. Complimentary searches in dijet final states probe lower masses up to 10 GeV~\cite{Sirunyan:2019sgo}. However the limits from these searches are weak. For \zprime masses below 100 GeV, the most relevant limits arise from low energy colliders, fixed target experiments and from electroweak precision tests. Among colliders, BaBar reaches the lowest \zprime mass of 0.05 GeV, up to 10 GeV of \zprime mass, the limit of \gprime is approximately constant at $10^{-4}$. For \zprime mass between 10 to 70 GeV, the strongest limits are placed by LHCb and they constrain \gprime < $10^{-3}$. For light \zprime below 1 GeV, there are stronger limits from fixed target experiments, however this region is not of primary interest to this work.  

\begin{figure}[t!]
\centering
\includegraphics[width=0.75\textwidth]{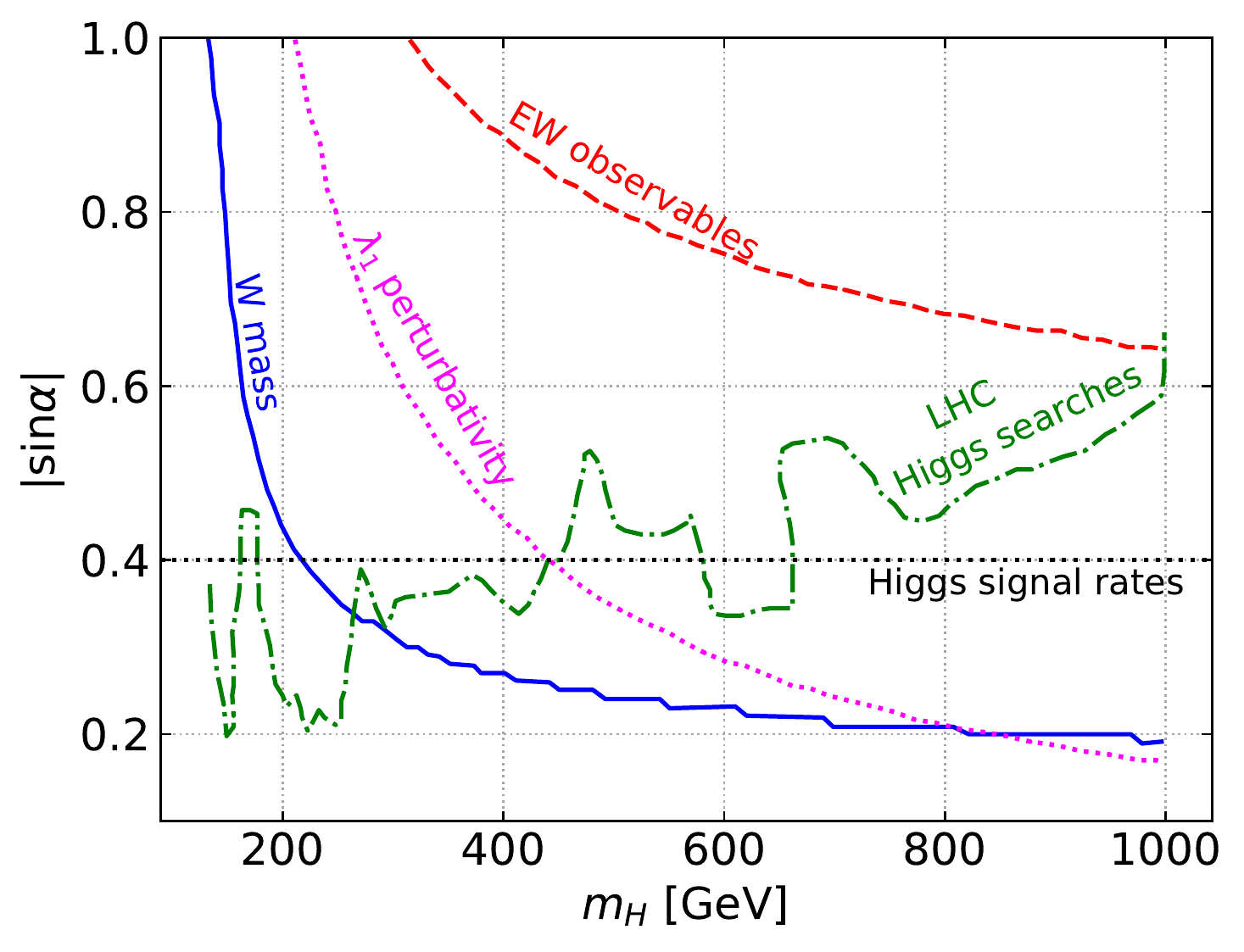}
\caption{Summary of constraints on the Higgs mixing angle $|\sin\alpha|$ as a function of the mass $m_H$ of the additional Higgs for a fixed value of $\tan\beta = v/\langle\chi\rangle = 0.1$. As indicated, the constraints arise from theoretical considerations of perturbativity (dotted magenta), indirect experimental constraints from corrections to the $W$ boson mass (solid blue), electroweak precision observables (dashed red), direct LHC searches for additional Higgs bosons (dash-dotted green) and LHC measurements of the Higgs signal rates (dotted black). The plot is adapted from~\cite{Ilnicka:2018def}.}
\label{fig:BSM_higgs}
\end{figure}
The other relevant sector for us is that of the Higgs. The singlet scalar $\chi$ and its mixing angle $\sin\alpha$ with the SM Higgs can be constrained in various ways. From theoretical consistency arguments, perturbativity, unitarity and vacuum stability requirements set limits on the quartic couplings of Higgs sector. The current constraints on the Higgs mixing angle $|\sin\alpha|$ as a function of the heavy Higgs mass and for a fixed value of $v/\langle\chi\rangle = 0.1$ is taken from Ref.~\cite{Ilnicka:2018def} and is summarised in Fig.~\ref{fig:BSM_higgs}. Direct limits from the Higgs signal strength measurements put a global upper limit of $|\sin\alpha|\lesssim 0.4$ \cite{Bechtle:2013xfa, Ilnicka:2018def, Bechtle:2014ewa, ATLAS:2016gld, ATLAS:2016oum, ATLAS:2016pkl, ATLAS:2016awy, ATLAS:2016nke, ATLAS:2016ldo, Sirunyan:2017exp, CMS:2016ixj}, independent of the mass of the heavy Higgs. The LHC searches for additional Higgs bosons tightly constrain the presence of extra Higgses with masses below $m_H = 300$~GeV. Above the mass of 300~GeV, the strongest limits are obtained by considering corrections to the $W$ mass and they limit the mixing angle at $|\sin\alpha| < 0.3$ for a heavy Higgs mass of 300~GeV. The constraint gets tighter as the heavy Higgs mass increases and in the limiting case of Higgs mass of 1~TeV, the limit approaches $|\sin\alpha| \lesssim 0.2$. For such large heavy Higgs masses however the perturbativity of the $\lambda$ coupling gives similar constraint on $|\sin\alpha|$. Consequently, we take the Higgs mixing up to its maximally allowed value $|\sin\alpha| = 0.3$ combining the allowed region near $m_{h_\chi}\approx 300$~GeV mainly from the strong limits from the $W$ boson mass and the direct Higgs searches. It is important to note that the limit on $|\sin\alpha|$ gets only mildly stronger for heavier Higgs masses. In the future, the limit on the Higgs mixing could be improved considerably to $\sin \alpha \lesssim 0.06$ at a lepton collider such as CEPC~\cite{Gu:2017ckc, CEPCStudyGroup:2018ghi} or the FCC-ee~\cite{Gu:2017ckc}. 

\subsection{\boldmath \zprime Decays}

%
\begin{figure}[t!]
	\centering
	\includegraphics[width=0.75\textwidth]{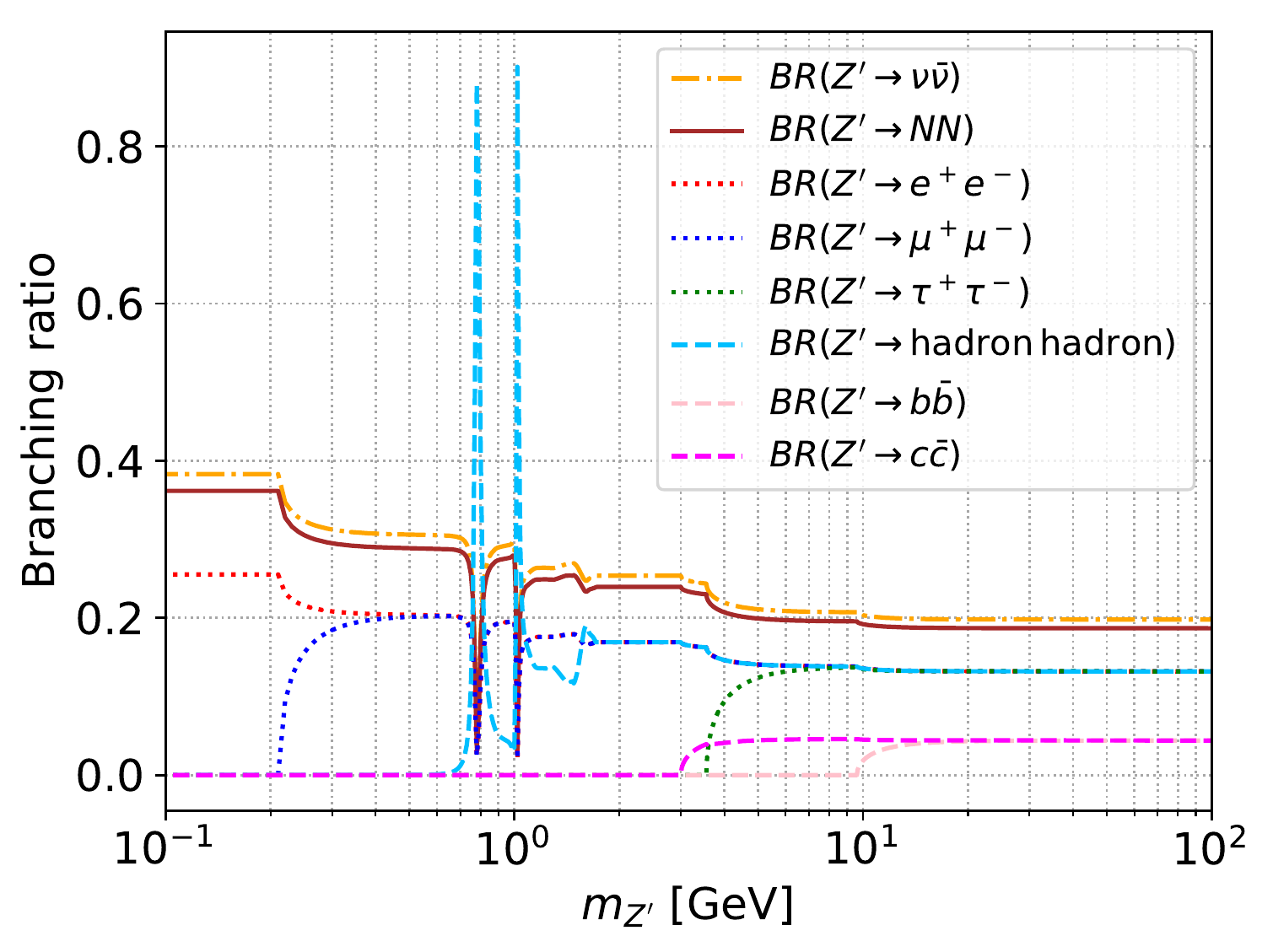}
	\caption{\zprime branching ratio to various final states as a function of \zprime mass. The mass of the three heavy degenerate neutrinos $N_i$ in the model is fixed to $m_N = m_{Z^{\prime}}/3$.}
	\label{fig:zp_decay_length}
\end{figure}
Another important quantity for this work is the \zprime decay length. The decay length in general is a function of the \zprime mass and the \gprime coupling. As \gprime decreases, it is possible for \zprime to obtain macroscopic decay lengths. The total decay width of $Z^{\prime}$ can be approximately expressed as
\begin{align}
\label{totaldecay}
	\Gamma(Z^{\prime})\approx \frac{23}{9}\frac{g^2_{B-L}}{4\pi} m_{Z^{\prime}},
\end{align}
for $m_{Z'} \gtrsim 2m_\mu$ and neglecting the effect of QCD resonances. It gives rise to an approximate $Z'$ proper decay length of
\begin{align}
\label{eq:l0}
	L_0 \approx 1~\text{mm}\cdot\left(\frac{10^{-6}}{g_{B-L}}\right)^2 
	    \cdot \left(\frac{1~\text{GeV}}{m_{Z'}}\right).
\end{align}
For small \zprime masses $m_{Z'} \lesssim 1$~GeV the leading order $Z'$ branching ratio computation as done by MadGraph may not be accurate and non-perturbative QCD effects become important. These effects are accounted for by scaling the branching ratio to the corresponding results obtained by the {\tt Darkcast} calculation. {\tt Darkcast} considers these effects by means of the vector meson dominance mechanism~\cite{Fujiwara:1984mp}. As the $B-L$ model in {\tt DarkCast} does not contain heavy neutrinos, three degenerate heavy neutrinos are also added to model accurately the branching ratio computation. 

In Fig.~\ref{fig:zp_decay_length}, we plot the \zprime branching ratios to the SM states and the heavy neutrinos. The model contains three heavy neutrinos which we choose to be degenerate at a mass of $m_N = m_{Z^\prime}/3$. This maximizes the branching ratio to neutrinos and minimizes that to muons, making it a conservative choice for our analysis, although the effect is in any case small. Beyond a \zprime mass of 1~GeV the branching ratios approximately remain constant with the exception of $\tau\tau$, $c\bar{c}$ and $b\bar{b}$ having thresholds at 3.0, 3.4 and 9.6~GeV, respectively. For \zprime masses below 1~GeV several thresholds due to QCD hadrons are visible. Effects due to loop corrections are accounted for as described before. Of particular importance for this work is the $BR(Z^\prime \to l^+l^-)$, which is approximately constant at 15\% per lepton species. 

\section{\boldmath \zprime Production Mechanisms}
\label{param_space}

\begin{figure}[t!]
\centering
\includegraphics[width=0.28\textwidth]{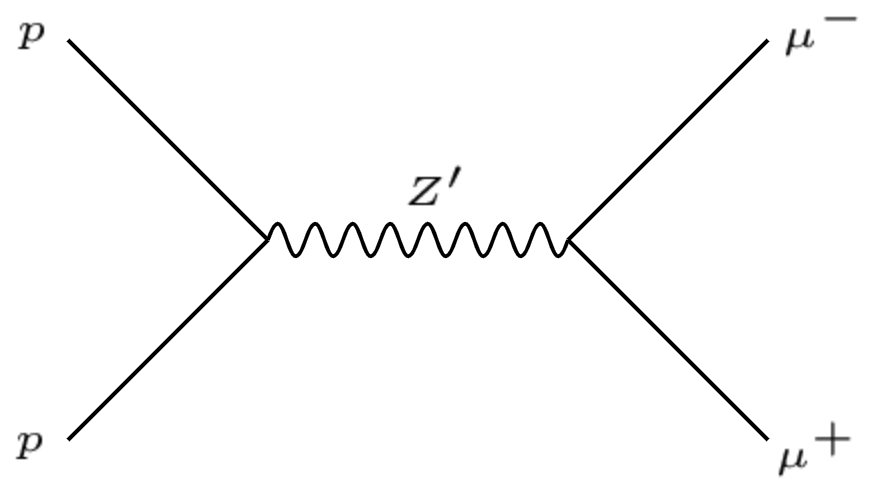}
\includegraphics[width=0.32\textwidth]{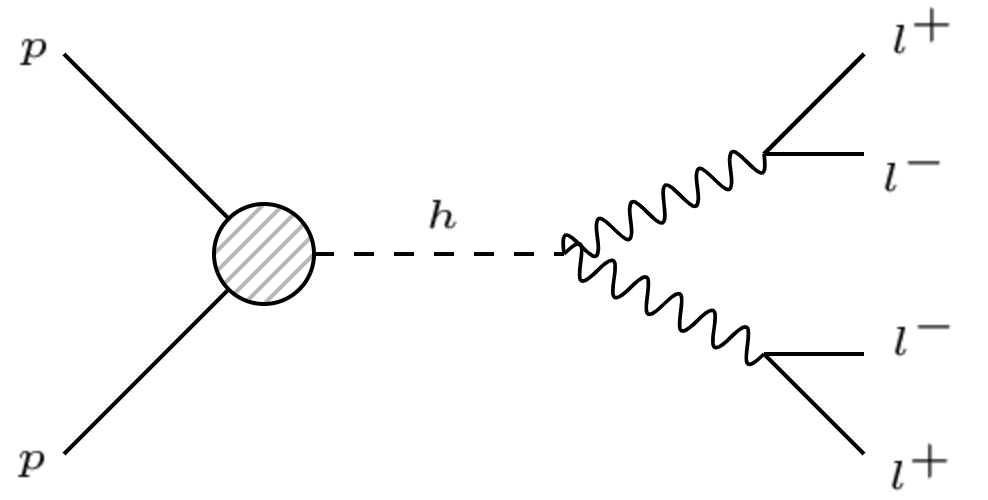}
\includegraphics[width=0.32\textwidth]{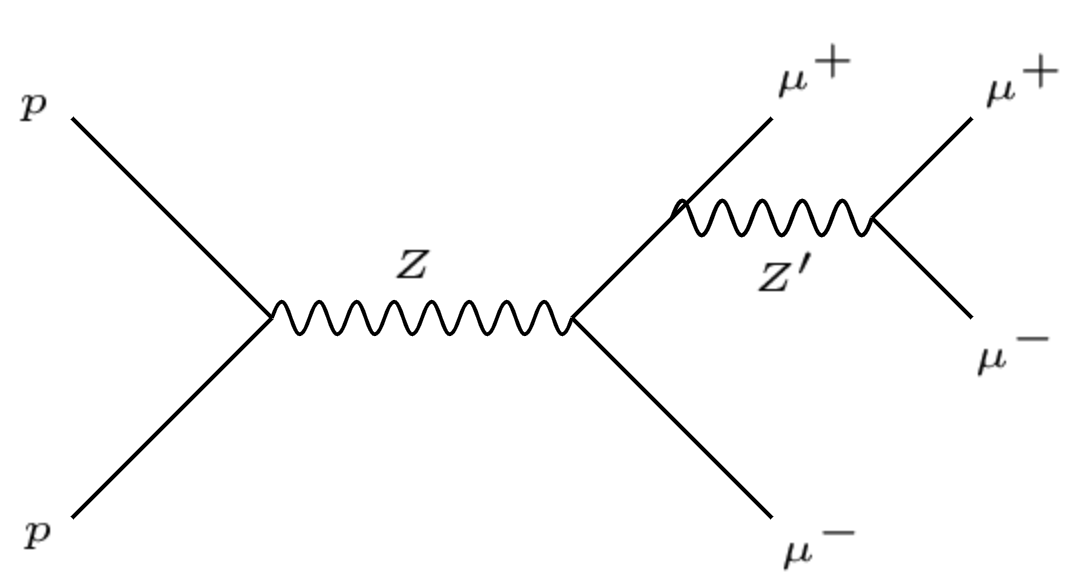}
\caption{Feynman diagrams of $Z'$ production modes considered in this paper: Drell-Yan $Z'$ $s$-channel $pp\to Z' \to \mu^+\mu^-$ (left), $Z'$ pair-production via SM Higgs $pp\to h \to Z'Z' \to 4\mu$ (center), and $Z'$ final state radiation (FSR) from SM $Z$ $pp\to Z \to Z'\mu^+\mu^- \to 4\mu$ (right).}
\label{fig:feynman}
\end{figure}

A potentially large Higgs mixing angle of order $\sin\alpha \approx 0.3$ implies that it is possible to produce the \zprime via decays of the SM Higgs. This presents an alternative and interesting possibility to probe the $Z'$ which is usually searched for via $s$-channel Drell-Yan production mode, cf. Fig.~\ref{fig:feynman}~(left). In the Higgs mediated case, Fig.~\ref{fig:feynman}~(center), the production is through a different vertex driven be the Higgs mixing $\sin\alpha$ but $m_{Z'}$ is restricted to $m_{Z'} < m_h/2\approx 62$~GeV. Alternatively, it is also possible to produce the \zprime via final state radiation in the Drell-Yan $Z$ production at the LHC, cf. Fig.~\ref{fig:feynman}~(right). 
 
With these observations, we will concentrate on three distinct processes at the LHC: Drell-Yan \zprime production, $Z'$ pair-production through SM Higgs and final state radiation (FSR) of \zprime in the Drell-Yan production of a SM $Z$. We concentrate on leptonic final states. In particular, we analyse the reach of following LHC analyses for the given final states 
\begin{itemize}
	\item $pp \to Z'\to \mu^+\mu^-$~\cite{CMS-PAS-EXO-19-018},
	\item $pp \to h \to Z' Z' \to 4\mu$~\cite{Sirunyan:2018mgs}, $\to 4\ell$~\cite{Aaboud:2018fvk},
	\item $pp \to Z \to \mu^+\mu^- Z' \to 4\mu$~\cite{Sirunyan:2018nnz}.
\end{itemize}

Apart from the above processes, it is also possible to search for \zprime via associated production with $W$/$Z$ or a jet. These processes will however yield a smaller production cross section compared to the ones listed above. Therefore, we will not explicitly consider them in this work. The associated production is accounted for in the $s$-channel production of \zprime in the form of showering and hadronization. As mentioned before, numerous resonance search results in dijet final states are also available. Their limits are however weak and therefore, we do not consider hadronic searches in this work.

Before proceeding with a detailed collider analysis of the above processes, it is instructive to take a look at the production cross sections and branching ratios relevant in each channel.

\subsection{\boldmath Drell-Yan \zprime Production}
%
\begin{figure}[t!]
	\centering
	\includegraphics[width=0.49\textwidth]{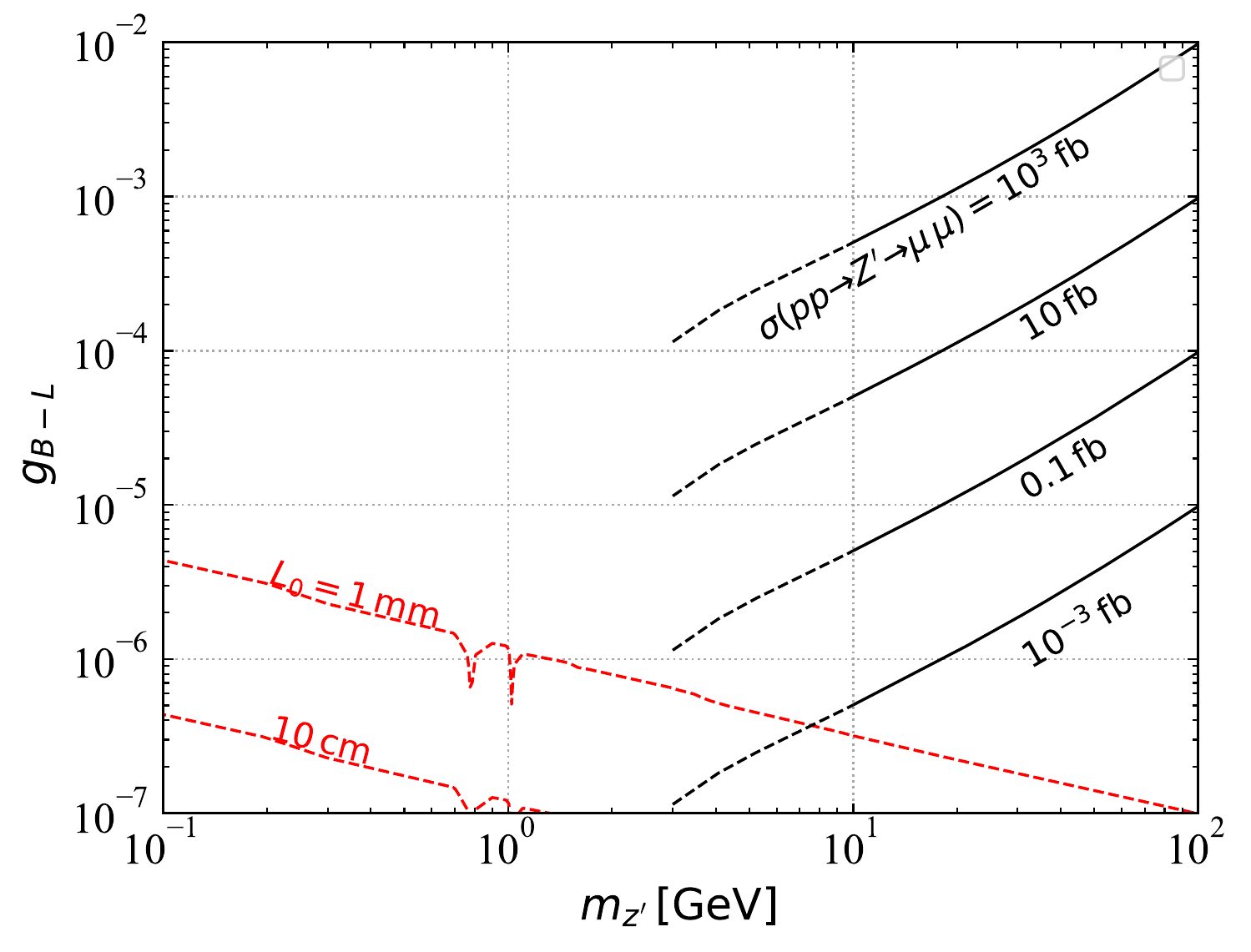}
	\includegraphics[width=0.49\textwidth]{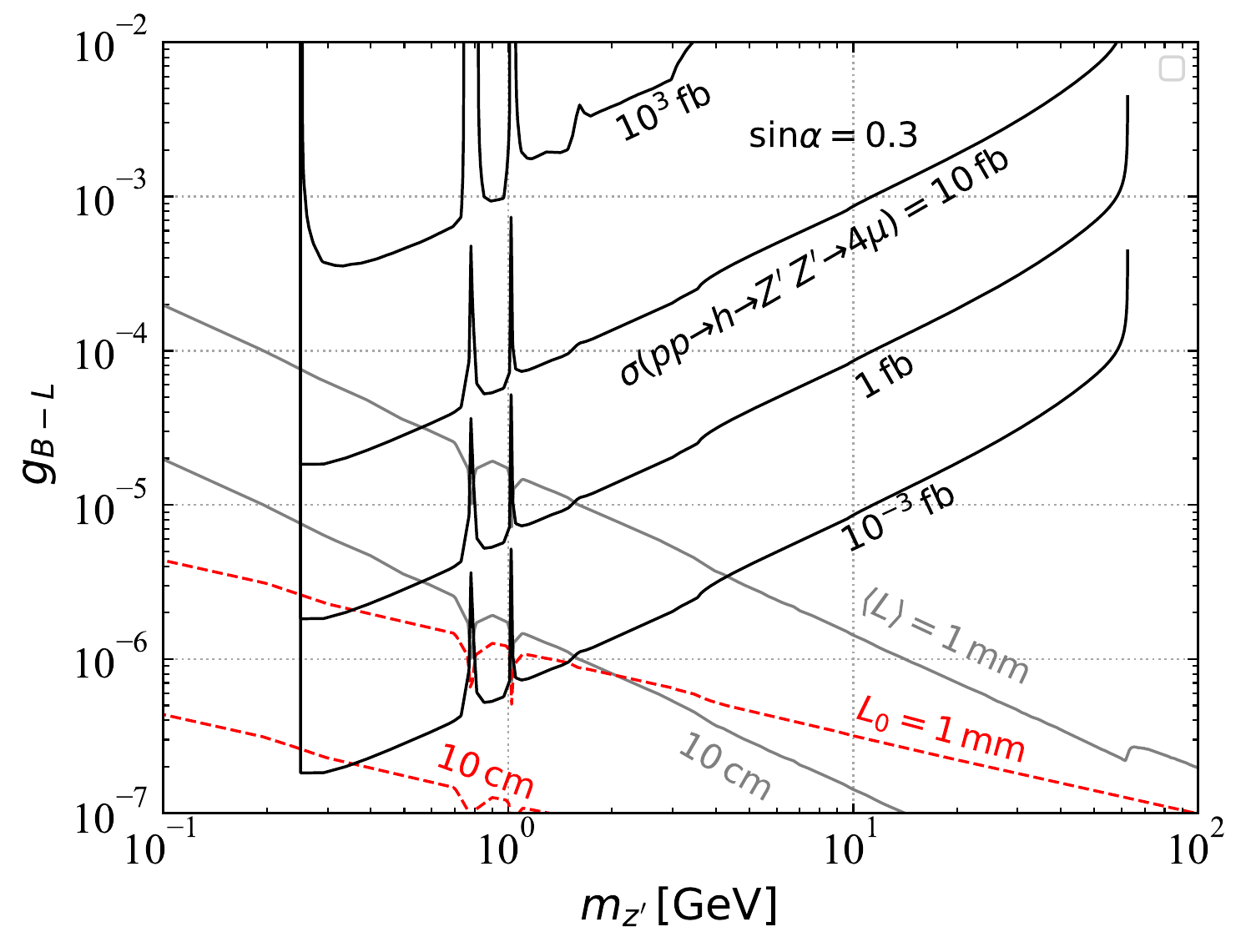}
	\includegraphics[width=0.49\textwidth]{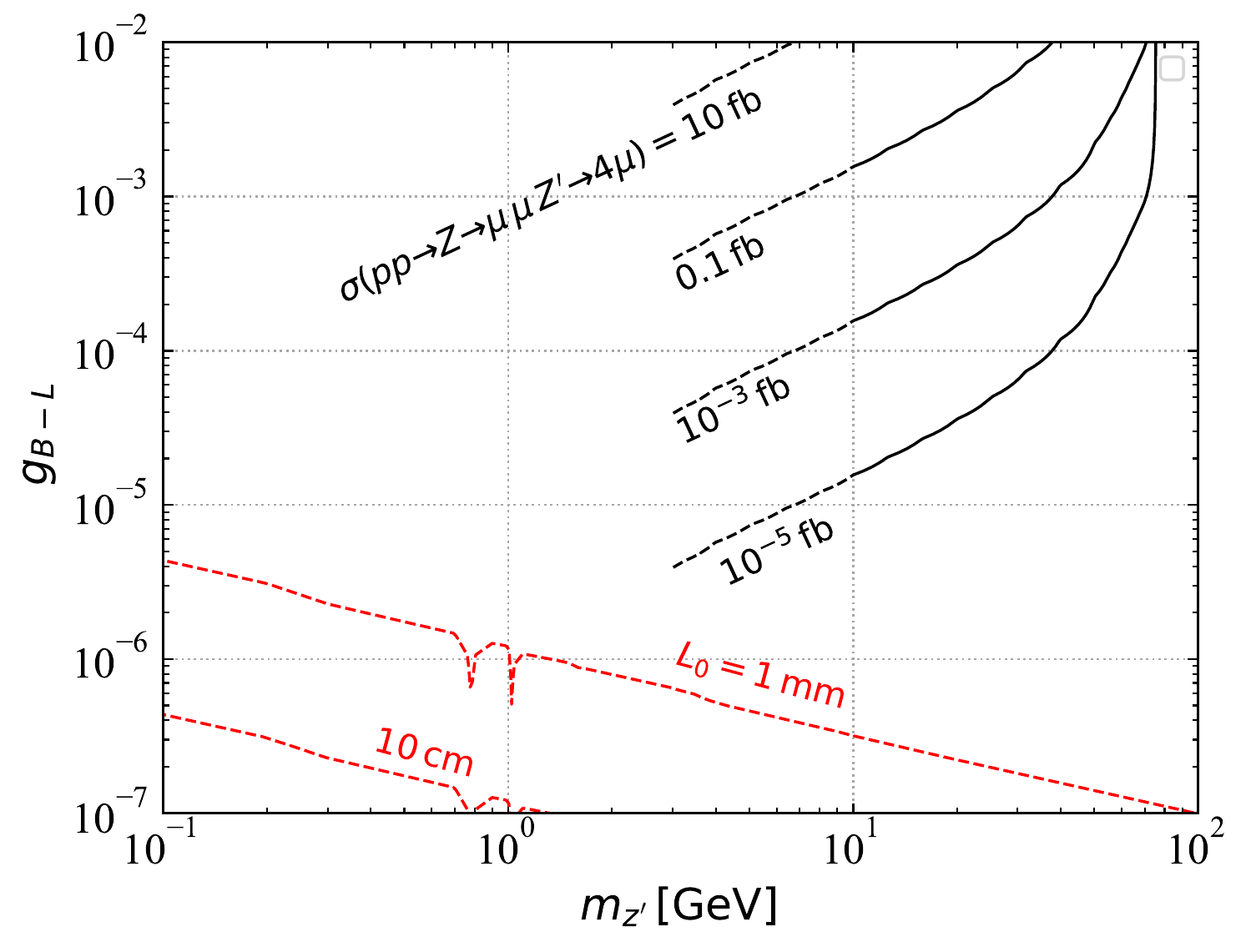}
	\caption{Cross section of Drell-Yan $Z'$ production, $pp\to Z' \to \mu^+\mu^-$ (top left), $Z'$ pair-production via Higgs, $pp\to h \to Z'Z' \to 4\mu$ (top right), and $Z$ production with final state radiation (FSR), $pp\to Z \to 2\mu + Z' \to 4\mu$ (bottom), as a function of the $Z'$ mass $m_{Z'}$ and the $U(1)_{B-L}$ gauge coupling $g_{B-L}$ (black solid curves). All cross section are at 13~TeV. In the upper right plot, the Higgs mixing is set to $\sin\alpha = 0.3$. The red lines represent the proper decay length $L_0$ of the \zprime as indicated whereas the blue lines in the upper right plot indicate the average \zprime lab frame displacement $\langle L \rangle$ as it is produced via the SM Higgs.}
	\label{fig:brcs}
\end{figure}
The $Z^\prime$ can be directly generated via $pp$ collisions at the LHC through $s$-channel Drell-Yan production. The cross section is a function of $g_{B-L}$ and $m_{Z^\prime}$. For $g_{B-L} = 10^{-3}$, the production cross section varies from several pb for light $Z^\prime$ around 10~GeV to $\mathcal{O}(100)$~fb for a \zprime mass of 100~GeV. Fig.~\ref{fig:brcs}~(top left) illustrates the dependence of the cross section on \gprime and \mzprime. The cross section falls by two orders of magnitude for every order of magnitude change in \gprime. We also overlay a contour showing the proper \zprime displacement of 1~mm (dashed red line). It is clear that it will be difficult to probe large regions of displaced \zprime via Drell-Yan production as the \zprime cross section becomes very small for small \gprime. The cross section is only calculated for $m_{Z^\prime}$ greater than $\approx 10$~GeV to avoid non-perturbative effects. Although the \zprime production cross section is very large for small masses, it becomes increasingly challenging to search for such light mass \zprime at the LHC as both the signal and background event rates become too high. The limitations on the search due to trigger rates can be circumvented by means of data scouting techniques or trigger level analyses. We will below demonstrate the reach of a recent scouting analysis on low mass \zprime.

\subsection{\boldmath \zprime Pair-production via SM Higgs}
When $m_{Z'} < m_h/2$, the exotic gauge boson $Z'$ can be pair-produced via the SM-like Higgs $h$. We assume $m_{h_\chi} > m_h$ and thus the mostly exotic Higgs $h_\chi$ does not play a role in the process. If the Higgs mixing angle is at its currently allowed value, $\sin\alpha \approx 0.3$, the process $pp \to h \to Z' Z'$ can produce $Z'$ efficiently,
\begin{align}
\label{cs}
	\sigma(pp\to h\to Z'Z') 
	&= \sigma(pp\to h) \times \text{BR}(h\to Z'Z') \nonumber\\
	&= \cos^2\alpha\times \sigma(pp\to h_\text{SM}) 
	   \frac{\Gamma(h\to Z'Z')}{\cos^2\alpha\,\Gamma(h_\text{SM}) 
		    +\Gamma(h\to Z'Z')},
\end{align}
where $\sigma(pp\to h_\text{SM})\approx 44\pm 4$~pb is the pure SM Higgs production cross section at 13~TeV \cite{deFlorian:2016spz} and $\Gamma(h_\text{SM}) \approx 4$~MeV is the total Higgs width in the SM \cite{Tanabashi:2018oca}. In Eq.~\eqref{cs}, we neglect the small partial width of the Higgs decaying to heavy neutrinos, $\Gamma(h\to NN)$, when calculating the total width \cite{Deppisch:2018eth}. The partial decay width to $Z'Z'$ is in our model given by
\begin{align}
\label{eqn:gammah1zpzp}
	\Gamma(h\to Z'Z') 
	&= \frac{{3\,g^2_{B-L}\sin^2\alpha}}{8\pi m_{Z'}^2} m^3_h
	   \sqrt{1 - \left(\frac{2m_{Z^\prime}}{m_h}\right)^2}
	   \left(1 - 4\left(\frac{m_{Z^\prime}}{m_h}\right)^2 
	           + 12\left(\frac{m_{Z^\prime}}{m_h}\right)^4\right).
\end{align}

The cross section in Eq.~\eqref{cs} is shown in Fig.~\ref{fig:brcs}~(top right) as a function of $m_{Z'}$ and $g_{B-L}$, where the Higgs mixing is set to $\sin\alpha = 0.3$. Also superimposed are contours of constant proper $Z'$ decay length $L_0$ in the rest frame and the average decay length $\langle L \rangle$ in the lab frame at 1~mm and 10~cm. The proper and average lab frame displacements are very different for lighter $Z'$ due to the associated boost. It can also be seen that \zprime starts to be appreciably displaced with $\langle L \rangle = 1$~mm for this production mode for $g_{B-L} \approx 10^{-4}$ when \mzprime is less than a GeV. As $m_{Z'}$ increases small values of \gprime are required to gain the same displacement. At this point however, the Higgs mediated \zprime production cross section is very small.

\begin{figure}[t!]
\centering
\includegraphics[width=0.50\textwidth]{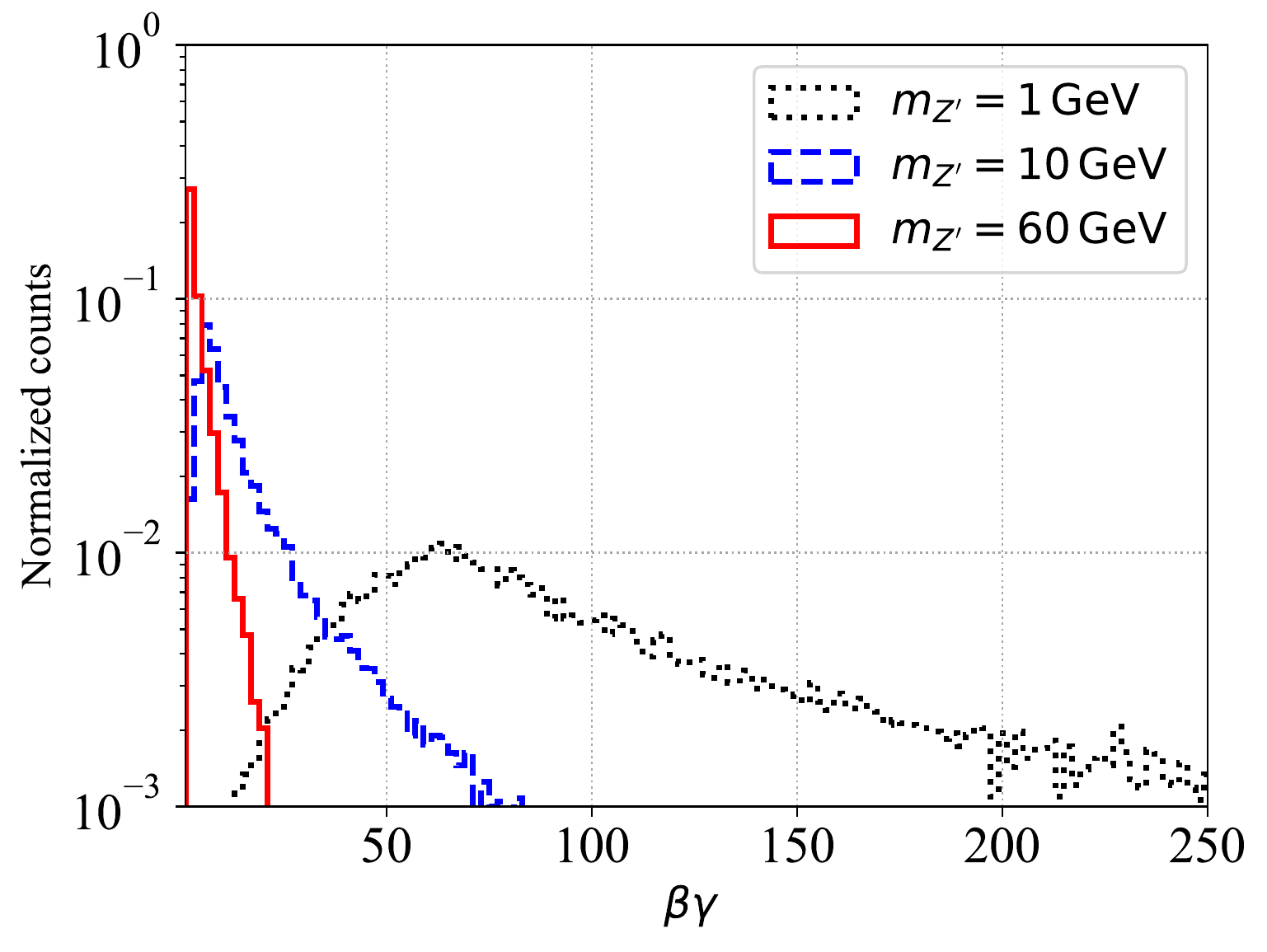}
\caption{Truth level distribution of the $Z'$ boost factor $\beta\gamma$ in  \zprime pair-production via SM Higgs, $pp\to h\to Z'Z'$.}
\label{fig:zp_disp}
\end{figure}
To better understand the \zprime boost and corresponding \zprime lab decay length, we show in Fig.~\ref{fig:zp_disp} the distribution for the boost factor $\beta\gamma = |\mathbf{p}_{Z'}|/m_{Z'}$ for different $m_{Z'}$ ranging from 1 to 60~GeV. For a light \zprime with $m_{Z'} = 1$~GeV, typical boost factors can reach beyond 100 but this decreases steadily for heavier $Z'$ and for $m_{Z'} = 60$~GeV, $\langle\beta\gamma\rangle$ is of order unity. For large $\beta\gamma$ the lab frame displacement can be very large even if the proper decay length is microscopic. At the LHC however, it will be difficult to probe a large region of the parameter space as the $hZ'Z'$ coupling is also proportional to \gprime. Therefore, a small value of \gprime leads to a small production cross section. As we will discuss later, this interplay of boost, corresponding lab frame decay length and suppression of production cross section leads to interesting results. 

\begin{figure} [t!]
	\centering
	\includegraphics[width=0.49\textwidth]{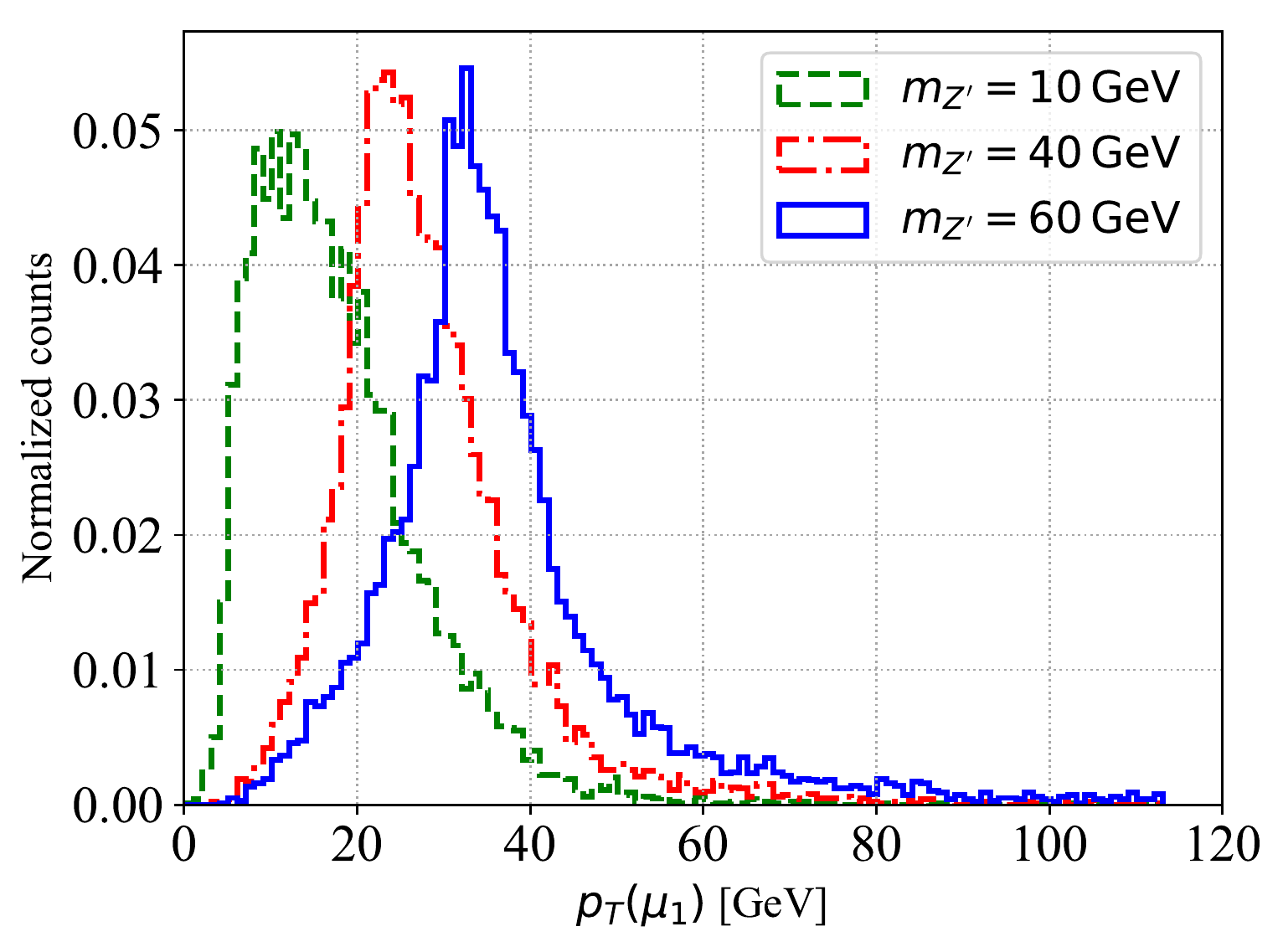}
	\includegraphics[width=0.49\textwidth]{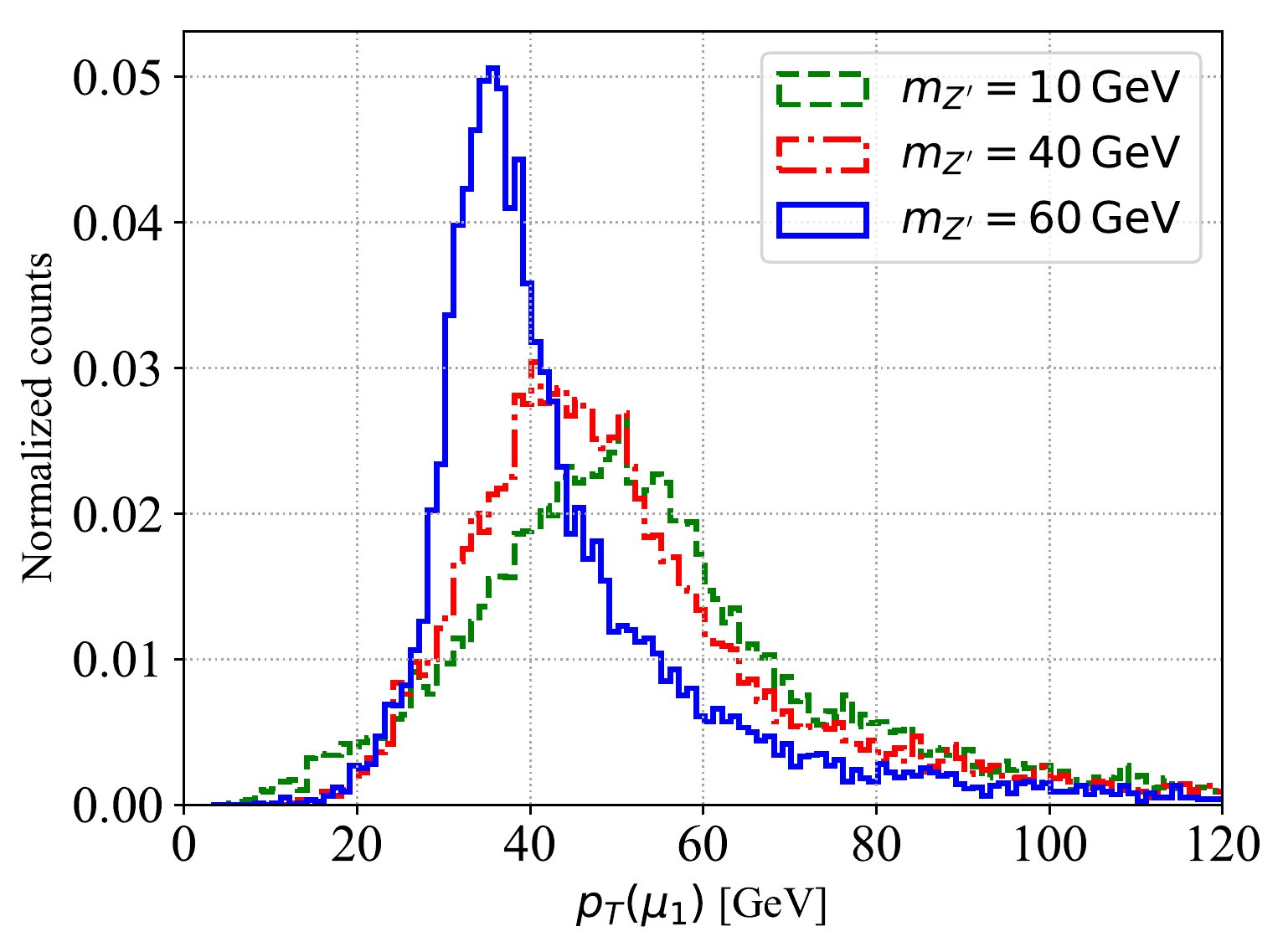}
	\caption{Truth level distribution of the transverse momentum of the leading muon for the Higgs production mode $pp\to h\to Z'Z'\to 4\mu$ (left) and the FSR mode $pp\to Z\to Z'\mu^+\mu^-\to 4\mu$ (right), for three different values of $m_{Z'}$.}
	\label{fig:kinematics}
\end{figure}
Along with the estimates of the total cross section it is imperative to gain an understanding of the broad kinematics of the processes we are considering. To this extent, in Fig.~\ref{fig:kinematics}~(left) we plot the $p_T$ distribution of the final state leading muon for Higgs mediated \zprime production. The $p_T$ of the leading muon increases as the mass of \zprime increases. This is to be expected as the Higgs is produced almost at rest and the muon $p_T$ is controlled by the \zprime mass. 

\subsection{\boldmath \zprime Final State Radiation from $Z$ Production}
In the context of LHC analyses, the production of a $Z'$ radiating from a lepton is particularly useful to explore in models, such that based on $L_\mu - L_\tau$, where it is not possible to produce the associated \zprime from quark annihilation. In case of the $B-L$ \zprime this is not really necessary, however, for completeness we discuss this process and show the total cross section as well as later determine the resulting constraints from this channel. As shown in Fig.~\ref{fig:brcs}~(bottom), the overall production cross section of this process is rather small as an emission of a massive particle from final state muon requires the muon to be off-shell and hence it is phase space suppressed. The cross section is only calculated for $m_{Z^\prime}$ greater than 10~GeV to avoid non-perturbative effects. The cross section attains a maximal value of 1~fb in the considered parameter space for $g_{B-L} < 10^{-3}$. In the leading muon $p_T$ distribution~\footnote{No interference effects with the SM $Z$ are taken into account at this point.}, Fig.~\ref{fig:kinematics}~(right) for this process, the dependence on $m_{Z'}$ is reversed to that for the Higgs production. This is understood because the overall energy-momentum of the process is conserved. Therefore, to produce more and more massive gauge boson in the final state, the muons are required to be softer.

\section{Recasting Procedure}
\label{recast}

In this section, we explain our recasting procedures for the existing ATLAS/CMS searches so that we can apply them to the $B-L$ model considered here. We either exploit model-independent limits given by the collaborations or we use event simulations to calculate the corresponding fiducial cross section in the $B-L$ model which we then compare with the experimental limit. 

We use the Universal FeynRules Output (UFO)~\cite{Degrande:2011ua} for the $B-L$ model with next-to-leading order (NLO) QCD production, developed in Ref.~\cite{Deppisch:2018eth}, in combination with the Monte Carlo event generator {\tt MadGraph5aMC$@$NLO} -v2.6.3~\cite{Alwall:2014hca} at parton level. For every signal sample, we generate $10^4$ signal events. We then pass the generated parton level events to {\tt PYTHIA} v8.235~\cite{Sjostrand:2014zea} which handles the initial and final state parton showering, hadronization and heavy hadron decays. We do not simulate detector effects. Individual analysis efficiencies as described later are taken into account in order to obtain results. The analysis results we consider here either include fiducial cross sections reported in certain part of the phase space without detector effects, or experimental efficiencies which can be applied to hadronized events. We therefore do not compromise on the accuracy of our results due to the absence of a detector simulation. For the Higgs-mediated mode we use the NLO capabilities of our model to simulate Higgs production via gluon-gluon fusion. 

\paragraph{\boldmath \zprime Pair-production via SM Higgs at CMS (CMS $h\to 4\mu$)}
In Ref.~\cite{Sirunyan:2018mgs}, the CMS collaboration reported on a search for the pair-production of new light bosons decaying into muons at $\sqrt{s} = 13$~TeV with an integrated luminosity of 35.9~fb$^{-1}$. The search was optimised for prompt exotic boson decays as well as those with moderate displacements. As results of the search, upper limits on the signal cross sections in prompt final state are presented for neutral boson masses between 0.25 and 3.55~GeV, or approximately $2m_\mu$ and 2$m_\tau$. However, the analysis is valid for a di-muon invariant mass up to $\approx 9$~GeV. We therefore reimplement the analysis and derive limits for \mzprime up to 8.5~GeV. In order to achieve this, we use model-independent upper limits on the signal cross sections presented in the analysis. To derive the theory predictions, the analysis reports that the detector efficiency $\epsilon_\text{detector}$ is almost independent of the signal model. This has been demonstrated in the analysis by taking the ratio of the generator level acceptance $\alpha_\text{gen}$ with the total efficiency for several signal samples. This factor, $\epsilon_\text{detector} = \alpha_\text{gen}/\epsilon_\text{total}$ is approximately constant and reported to be 60\%. In order to assist simulating $\alpha_\text{gen}$, the analysis advocates applying the cuts
\begin{align}
	p_T(\mu_1) &> 17~\text{GeV},\quad |\eta| < 0.9,\quad 
	\text{for the leading muon} \nonumber\\
	p_T(\mu) &> \phantom{0}8~\text{GeV},\quad |\eta| < 2.4,\quad 
	\text{for the other three muons}.
\end{align}
In addition, transverse, $L_{xy}$, and lateral, $L_z$, displacements of each muon from the interaction point are required to be 
\begin{align}
 	L_{xy} < 9.8~\text{cm},~~L_z < 46.5~\text{cm},
\end{align}
so, in fact, the selection criteria include scenarios where the $Z'$ can be appreciably long-lived with $L_0 \lesssim 10$~cm.

We have verified the reported $\alpha_\text{gen}$ by producing a sample of SM Higgs decaying to light $Z^\prime$ of 1~GeV and applying the cuts as reported above.  For our model, the cuts on the $L_{xy}$ or $L_z$ are not relevant in most of the parameter space as the $Z'$ is not long-lived. Nevertheless, for light $Z'$ with small $g_{B-L}$ the decay length in the lab frame can be macroscopic. For example, for $m_{Z'} = 0.25$~GeV and $g_{B-L} = 10^{-6}$ the $Z'$ proper decay length is $L_0 \sim$~cm. Accounting for an average Lorentz boost factor of about 100, the average decay length $\langle L\rangle$ can be as large as a meter. Therefore, the cuts become relevant for a small region in parameter space for our analysis.

The estimated background for this search is reported as $9.90\pm 1.24\,\text{stat} \pm 1.84\,\text{syst}$ events for 35.9 fb$^{-1}$~\cite{Sirunyan:2018mgs}. The 95~\% confidence level limit on the signal event rate can be derived from $\chi^2 = S^2/B$ > 3.84~\cite{Tanabashi:2018oca}. We have however used the model-independent limits given in Ref.~\cite{Sirunyan:2018mgs}. The current sensitivity in our parameter space is obtained by requiring $\sigma_\text{model} < \sigma^\text{upper limit}_\text{exp}$; the limits from $\chi^2$ are however very similar. 

Furthermore, we also compute the reach of this analysis for High Luminosity LHC (HL-LHC) regime with 3000 fb$^{-1}$ luminosity. As the analysis is sensitive to low mass \zprime, where the \zprime can obtain macroscopic displacement, we implement the analysis cuts as described before, and compute the HL-LHC reach with $\chi^2$ analysis.

\paragraph{\boldmath \zprime Pair-production via SM Higgs at ATLAS (ATLAS $h\to 4\,l$)}
A corresponding ATLAS analysis~\cite{Aaboud:2018fvk} reports upper limits (U.L.) on the signal strength for pair production of light exotic bosons through decays of the SM Higgs at $\sqrt{s} = 13$~TeV with an integrated luminosity of 36.1~fb$^{-1}$. The analysis searches for light bosons decaying to either pair of electrons or muons, hence it searches for either $4e, 4\mu$ or $2e2\mu$ final states. These signal strength limits are given for the light boson decaying promptly between a mass range of 1 to 60 GeV, with the SM QCD resonance regions removed. The signal strength is the ratio of model specific Higgs production cross section with the SM Higgs production cross section. In our model this ratio is $\cos^2\alpha$. The limits given on the signal strength hence convert to a limit on the Higgs to \zprime branching ratio. We compare this to the theoretical prediction BR$^{th}(h\to Z'Z')$,
\begin{align}
 	 \left(\frac{\sigma(h)}{\sigma(h_\text{SM})} \times \text{BR}(h\to Z'Z')\right)^\text{U.L}
 	 = \cos^2\alpha\times \text{BR}^{\text{U.L.}}(h\to Z'Z')
 	 = \text{BR}^\text{th}(h\to Z'Z').
\end{align}
Unlike in the case of the CMS analysis, here we directly use the limits on the signal strength. These are derived under the assumption of a promptly decaying \zprime. As discussed in Sec.~\ref{param_space}, probing small values of \gprime -- \zprime can lead to displaced vertices. Therefore, one should be careful while interpreting the limits of analyses which assume prompt final states only, as is the case here. We consider \zprime to be prompt when their lab frame displacement $\langle L \rangle$ is less than 1~mm. This is fixed by inserting a prompt efficiency function $\epsilon_\text{prompt}\approx 1-\exp(-\text{1~mm}/L)$. We use the same function for the HL-LHC projections as well.

\paragraph{\boldmath \zprime Final State Radiation from $Z$ Production at CMS (CMS FSR)}
The CMS analysis~\cite{Sirunyan:2018nnz} reports on the search for an excess in the $4\mu$ final state when the \zprime is radiated in the final state as $pp\to Z\to Z'\mu^+\mu^-\to 4\mu$ at the LHC with $\sqrt{s} = 13$~TeV with an integrated luminosity of 77.3~fb$^{-1}$. It considers \zprime mass range between 5 to 70~GeV. The analysis selects events with isolated muons. At least two muons are required to have $p_T > 20 \rm{GeV}$ and at least one muon should have $p_T > 10 \rm{GeV}$. A resonance search is then performed in pairs of oppositely charged muons. As the limits on the couplings between the two models can be easily converted from one to another, we do not perform any special simulation.   Instead we use the limits on $L_\mu - L_\tau$ coupling $g_{L_\mu - L_\tau}$ as given by CMS. An equivalent $g_{L_\mu - L_\tau}$ can be related to $g_{B-L}$ through $\sigma_\text{limit} \propto g_{B-L}^2 \times \text{BR}(Z^\prime\to\mu^+ \mu^-)_{B-L}$ = $g_{L_\mu - L_\tau}^2 \text{BR}(Z^\prime\to\mu^+ \mu^-)_{L_\mu - L_\tau}$,
\begin{align}
	g_{B-L}^2 = \frac{g_{L_\mu - L_\tau}^2}{3\times\text{BR}(Z^\prime\to\mu^+ \mu^-)_{B-L}}
\end{align}
because $\text{BR}(Z^\prime\to\mu^+ \mu^-)_{L_\mu - L_\tau} = 1/3$ \cite{Sirunyan:2018nnz}. For the HL-LHC regime, we rescale our limits. As the \zprime masses considered here are more than a few GeV, no special consideration for macroscopic decay lengths are given.

\paragraph{\boldmath Low mass $Z'$ Resonance Search at CMS (CMS dilepton)}
Finally, we include the most recent search for a narrow low mass resonance in the dimuon final state by CMS collaboration~\cite{CMS-PAS-EXO-19-018}. This $\sqrt{s} = 13$~TeV analysis uses 96.6~fb$^{-1}$ of data for a scouting search for a resonance between 11 to 45~GeV and the full 137~fb$^{-1}$ Run-II reconstructed level dataset for a resonance search between 45 and 200~GeV. The analysis overcomes the traditional limitations for dilepton resonance search in low mass region by making use of the data scouting technique. The technique corresponds to the use of physics objects reconstructed online during data taking to perform searches and measurements. This allows for reaching low mass resonances which are otherwise difficult to search for. 

The analysis interprets the results in a dark photon model and gives upper limits on the kinetic mixing $\epsilon$ as a function of dark photon mass $m_{Z_D}$. The limits are given in the mass range from 11.5 GeV up to 200 GeV $Z_D$ masses. The kinetic mixing parameter $\epsilon$ is related to $g_{B-L}$ by $g_{B-L} = e\,\epsilon$. 

The dimuon resonance analysis is applicable to a wide range of signal models. Therefore, it is possible to constrain not just the resonance production of \zprime but the Higgs mediated \zprime production as well. The cross section for the Higgs mediated \zprime production however is much smaller compared to the direct \zprime production for the same \gprime coupling. This is because the Higgs coupling to \zprime is suppressed by both \gprime and $\sin\alpha$. We therefore do not take into account the Higgs mediated process when computing the limits on \gprime.

\section{Results}
\label{results}

Using the above procedures for each of the existing searches, we determine the upper 95\% confidence level limits on the $U(1)_{B-L}$ gauge coupling as a function of the $Z'$ mass $m_{Z'}$. Unless stated otherwise we assume $\sin\alpha = 0.3$ for the mixing angle between the SM Higgs and the exotic scalar $\chi$ responsible for breaking the $U(1)_{B-L}$ symmetry. This assumption is of course crucial for the $Z'$ pair-production via the SM Higgs; for smaller values of $\sin\alpha$, the production rate is accordingly reduced and the limit on $g_{B-L}$ is weakened.  

\begin{figure}[t!]
\centering
\includegraphics[width=0.8\textwidth]{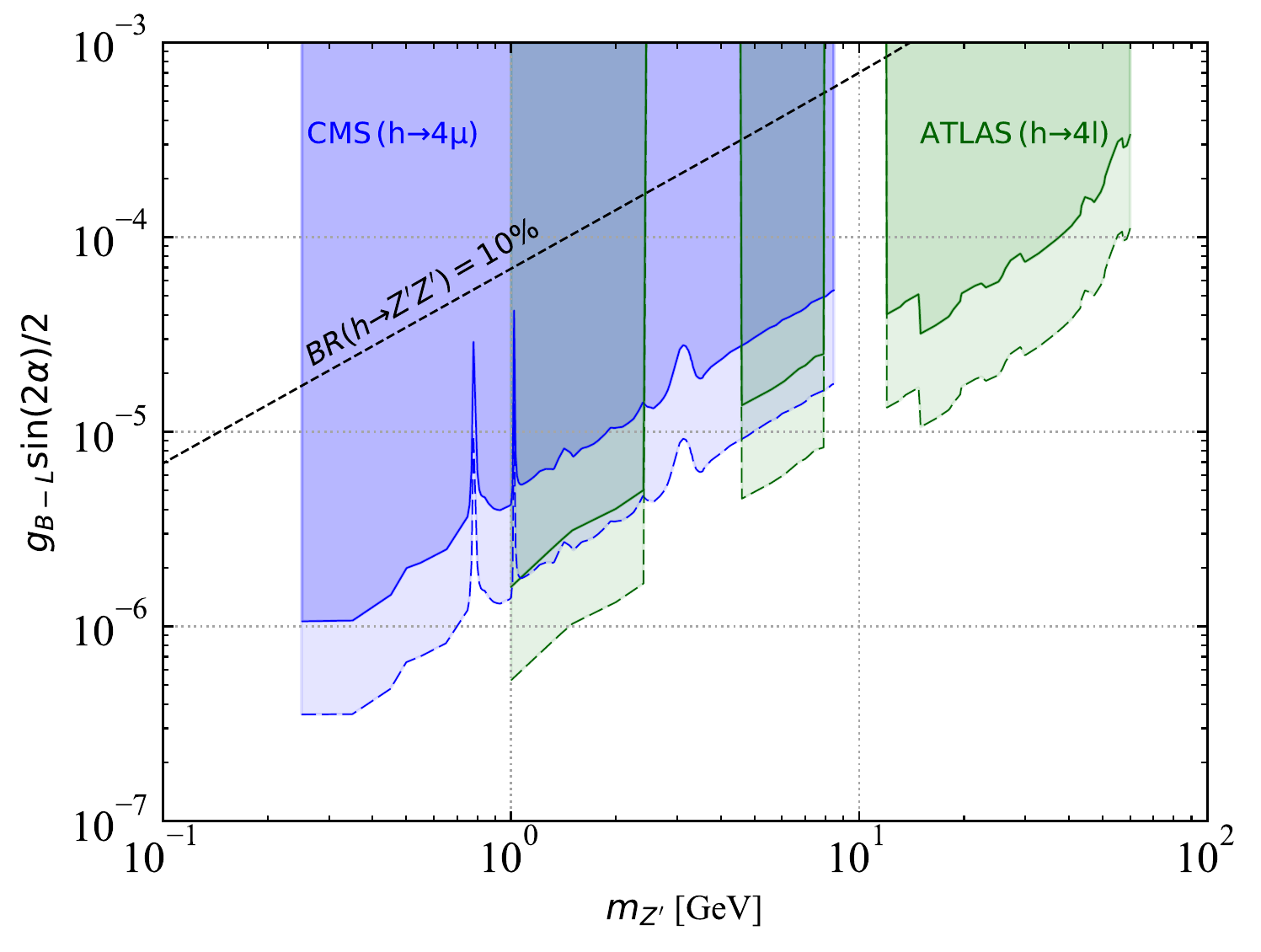}
\caption{Constraints on the effective coupling $g_{B-L}\sin(2\alpha)/2$ as a function of the $Z'$ mass $m_{Z'}$ derived from the CMS $h\to 4\mu$ \cite{Sirunyan:2018mgs} and ATLAS $h\to 4\ell $ \cite{Aaboud:2018fvk} searches. The dark coloured regions are excluded by current data whereas the light coloured regions indicate the improvement expected by rescaling to a luminosity of 3,000~fb$^{-1}$. Also indicated is the contour for constant branching ratio BR$(h\to Z'Z') = 10$\%.}
\label{fig:effective_coupling}
\end{figure}
More specifically, the Higgs production cross section effectively depends on the combination $g_{B-L}\sin(2\alpha)/2$, cf. Eq.~\eqref{cs}. In Fig.~\ref{fig:effective_coupling} we show the constraints from the analyses considered in this work on this parameter as a function of $m_{Z'}$. The CMS $h\to4\mu$ constraints span an $m_{Z'}$ mass range between 0.25 and 8.5~GeV, while the ATLAS $h\to4\ell$ cover the range between 1 to 60~GeV with two gaps between 2~GeV < \mzprime < 5~GeV and 8~GeV < \mzprime < 10.5~GeV arising from the requirement to remove QCD resonances. On the other hand, the CMS analysis estimates this background identifying correlations between di-muon invariant mass pairs. The ATLAS limits are stronger than the CMS limits where available. As can be seen from the model independent limits presented by both ATLAS and CMS, the limits on fiducial cross sections are very similar. However the phase space in which the fiducial cross section is computed is very different. In order to demonstrate the effect of phase space we can estimate the generator level acceptance for the two analysis. For this, we implemented the acceptance cuts as given in the two analyses, we find that the acceptance for the CMS analysis is about 25\% while that for the ATLAS analysis is 50\%. We also show the projections of the improved sensitivity for the high-luminosity LHC with 3,000~fb$^{-1}$ using the lighter shaded regions delimited by a dashed curve. This projection assumes a simple scaling of signal and background with luminosity. If $h\to Z'Z'$ is the only non-Standard Model Higgs decay mode available, then it is constrained by Higgs to invisible branching ratio. Therefore, we have also overlaid the line corresponding to BR$(h\to Z'Z') = 10$\% which corresponds to the existing limits on Higgs to invisible branching ratio. It should be noted that strictly speaking this does not include dependence on  $\cos\alpha$, however it does depend on $\sin\alpha$ as discussed in Eq.~\eqref{eqn:gammah1zpzp}. Furthermore, it is also worth pointing out that if heavy neutrinos are lighter than $m_h/2$, they will also contribute to this invisible BR constraint. Here we neglect the SM Higgs decaying to heavy neutrinos.

This plot is particularly useful as there is a degeneracy between \gprime and $\sin\alpha$ which can only be broken by individually searching for the presence or absence of extended Higgs or gauge sectors at experiments. Using this plot, it is possible to rescale and obtain values of \gprime for any value of $\sin\alpha$ desired. For example, for $m_{Z'} = 1$~GeV, at $\sin\alpha = 0.3$, $g_{B-L} \sim 3\times 10^{-5}$ but if $\sin\alpha = 0.2$, the limit on \gprime will be $\approx 5\times 10^{-5}$. On the other hand if we saturate the existing limits on the gauge coupling, $g_{B-L} \lesssim 10^{-4}$, the corresponding constraint on the Higgs mixing is $\sin\alpha \lesssim 0.1$. It will be difficult to independently constrain $\sin\alpha$ to such a small value by such means as direct Higgs searches. It is however very important to remember that such a compensation between $g_{B-L}$ and $\sin\alpha$ is not applicable to arbitrary low values of \gprime. As \gprime decreases, the \zprime will be longer-lived. The analyses we considered however largely concern themselves with prompt decays. Therefore for smaller values of \gprime a simple scaling between $\sin\alpha$ and \gprime will not hold true.  

Apart from the two Higgs searches considered in this work, an ATLAS search in lepton jets final state, interesting for the low mass \zprime region, has been carried out at $\sqrt{s} = 8\,\rm{TeV}$, both in prompt~\cite{Aad:2015sms} and displaced final states~\cite{Aad:2014yea}. The prompt analysis is of particular interest as it targets both muon and electron jets. The displaced lepton jets analysis is not sensitive to our model as we will not have large signal cross sections and displaced \zprime at the same time for high mass region where this analysis operates. The interpretation of results in both these analysis has been done in the so called FRVZ models which has substantially different kinematics to the model considered in this work. A reinterpretation of the prompt ATLAS lepton jet analysis can be attempted, however since it does not improve on the existing limits from 13 TeV CMS analysis considered here, we do not consider it here. Recently, ATLAS updated their search with 13 TeV data~\cite{Aad:2019tua}, however this analysis in not included in our work. 

\begin{figure}[t!]
\centering
\includegraphics[width=0.8\textwidth]{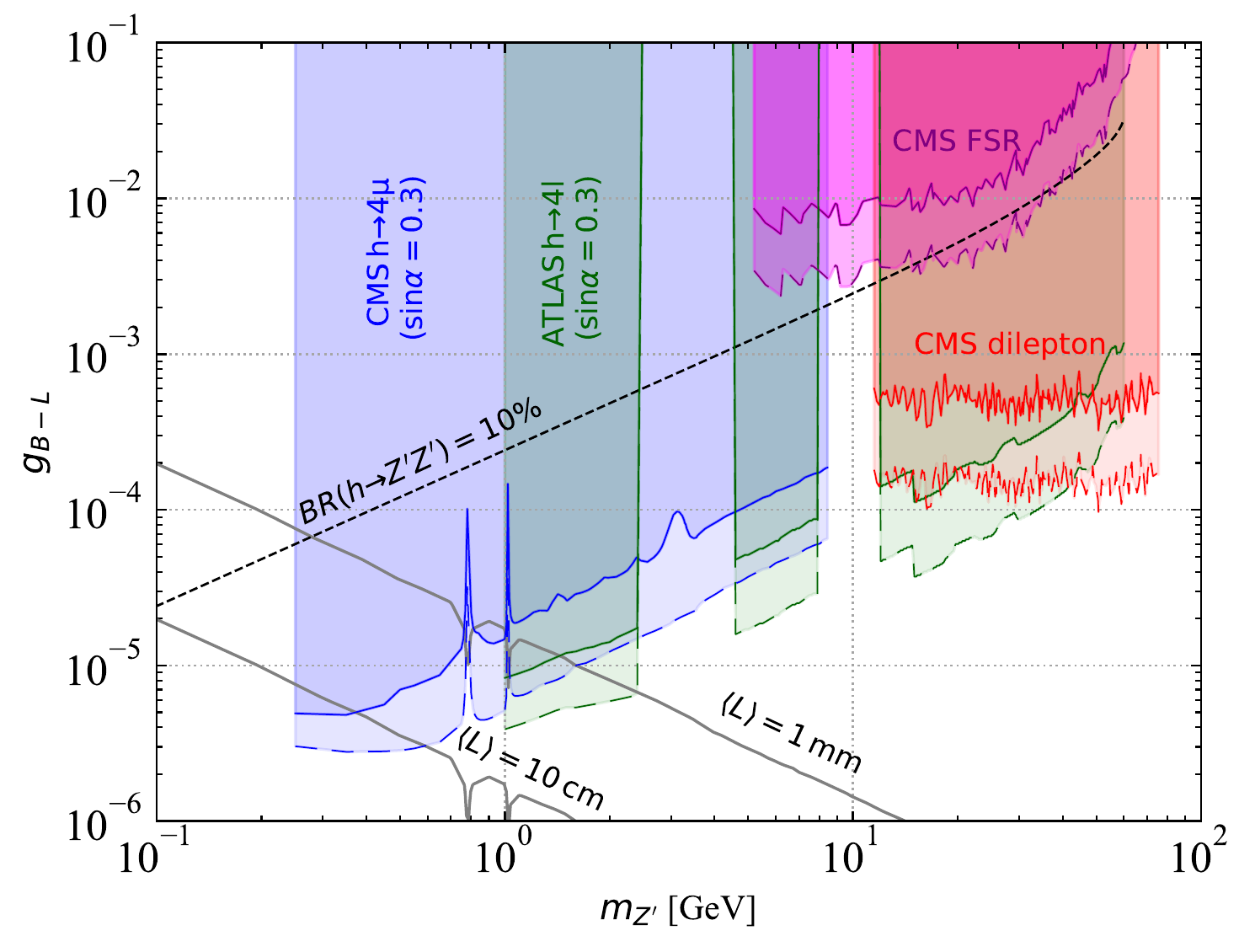}
\caption{Constraints on the $U(1)_{B-L}$ gauge coupling $g_{B-L}$ as a function of the $Z'$ mass $m_{Z'}$ derived from the CMS $h\to 4\mu$ \cite{Sirunyan:2018mgs}, ATLAS $h\to 4\ell $ \cite{Aaboud:2018fvk}, CMS FSR \cite{Sirunyan:2018nnz} and CMS dilepton \cite{CMS-PAS-EXO-19-018} searches. The dark coloured regions are excluded by current data whereas the light coloured regions indicate the improvement expected by rescaling to a luminosity of 3,000~fb$^{-1}$. Also indicated are the contours for constant branching ratio BR$(h\to Z'Z') = 10$~\% and constant average decay length $\langle L\rangle = 1$~mm, 10~cm, the latter applicable for the Higgs mode.}
\label{fig:limits_fixed_alpha}
\end{figure}
While Fig.~\ref{fig:effective_coupling} provides a concise summary of the Higgs mode searches, it does not accommodate search results which do not depend on $\sin\alpha$. In Fig.~\ref{fig:limits_fixed_alpha}, we instead show the limits for a fixed value of $\sin\alpha = 0.3$ but we additionally show constraints from the CMS FSR and CMS di-lepton searches. The constraints arising from the CMS FSR search leads to the weakest limit, $g_{B-L} \gtrsim 0.01$ for \mzprime in the range of 5 to 60~GeV. The 4$\mu$ final state arising due to the decays of SM Higgs to a pair of \zprime lead to strongest limits between \mzprime of 0.25 to 50 GeV. Beyond \mzprime = 50 GeV, the CMS dilepton analysis  leads to strongest limits up to 70 GeV. We also denote \zprime-\gprime values where average lab frame \zprime displacement of 1 mm and 10 cm is obtained when \zprime is produced in decays of h. Finally, for reference $BR(h\to Z' Z') =10\%$ is also overlaid.

For the lowest \zprime masses, the CMS $h\to4\mu$ analysis has the strongest limits. They constrain \zprime masses as low as 0.25 GeV and limit $g_{B-L}$ to $5\times 10^{-6}$.  These limits gradually decrease to $g_{B-L} = 1.8\times 10^{-4}$ for \mzprime = 8.5 GeV. As discussed in Sec.~\ref{param_space}, for $m_{Z'} < 1$~GeV, it is possible to gain a significant \zprime displacement. This will be relevant for the High Luminosity regime. In this region of parameter space a simple scaling is not applicable. We identify the region corresponding to $Z'$ lab decay length $\langle L\rangle$ of 1mm and 10cm (solid grey lines). A significant region of $m_{Z'}$ -- \gprime parameter space is below $\langle L\rangle = 1$~mm. As the CMS search allows for displacements up to 10~cm, it is perfectly safe to use the analysis in this region. The impact of 1~mm lab frame displacement is however more severe for the ATLAS search we consider as it only allows for prompt decays of the \zprime. We define the prompt region to be displacements less than 1~mm. This requirement has a mild effect on the analysis efficiency, however it is not visible in the final results. The HL-LHC reach for this analysis is correspondingly limited for low mass regions. Turning our attention to $\langle L\rangle = 10$~cm line, we see a similar picture emerge for the CMS search below $m_{Z'} < 0.5$~GeV. This is also understandable as the analysis allows for displacements up to 10~cm. In order to assess our reach for the High Luminosity reach in this region, we have taken into account the effect of displacement. This is reflected in the limits as the gain due to luminosity is much smaller in the displaced region $m_{Z'} \lesssim 0.5$~GeV compared to the prompt region $m_{Z'} \gtrsim 0.5$~GeV. The same can be seen for the ATLAS search.

Finally, we also show recent limits on dimuon final state resonance search using the data scouting technique as presented by the CMS collaboration~\cite{CMS-PAS-EXO-19-018}. This limit improves on the previous LHCb limit for a resonance search in the same final state from mass range of 11.5 GeV and presents competitive limits from the Higgs to 4 lepton final state in the mass range between 10 to 50 GeV. In the mass range of 50 to 70 GeV, this analysis has the best limits on \gprime.

Of particular interest is also the behaviour of limits from the FSR and dilepton final state against those from the Higgs mediated 4 lepton final state. The limits on the \gprime from the FSR and dilepton final states are approximately constant over a wide range of \zprime mass. The limits on \gprime coming from the Higgs mediated processes, however sharply degrade as \mzprime increases. This is because the 4 lepton final state cross section is dominantly controlled by the branching ratio of the SM-like Higgs decays to the $Z^\prime$ pairs, which depends on \gprime/\mzprime according to Eqs.~\eqref{eqn:gammah1zpzp} and~\eqref{cs}. 

Given our discussion so far, it is clear that it will be difficult to probe large \zprime dis\-place\-ments in the $B-L$ model unlike in the dark photon case.  However, the above discussion is built based on the assumption that we neglect the $Z-Z^\prime$ mixing of the $B-L$ model. If this mixing is opened complex interactions will be introduced as the $B-L$ sector can now couple to the SM particles via both the hypercharge portal and the $B-L$ charge. The relative strength of the two couplings \gprime and $\tilde g$ will then control the behaviour of the limits.

\section{Conclusions}
\label{conclusion}

In this work, we have considered the impact of LHC searches on the parameters of the minimal $B-L$ model for \zprime masses in the region $\approx 0.2$~GeV to 200~GeV. The minimal $B-L$ can be considered as simplest gauge realization to generate the light neutrino masses via a type-I seesaw mechanism and probing it will help in our understanding of neutrinos. The model presents a distinctly different phenomenology compared to the popular dark photon models. For example, unlike the dark photon models, the production and decay of $B-L$ \zprime are controlled by the same parameters, which limits the sensitivity of LHC searches due to rapidly falling cross sections. 

\begin{figure}[t!]
	\centering
	\includegraphics[width=1.0\textwidth]{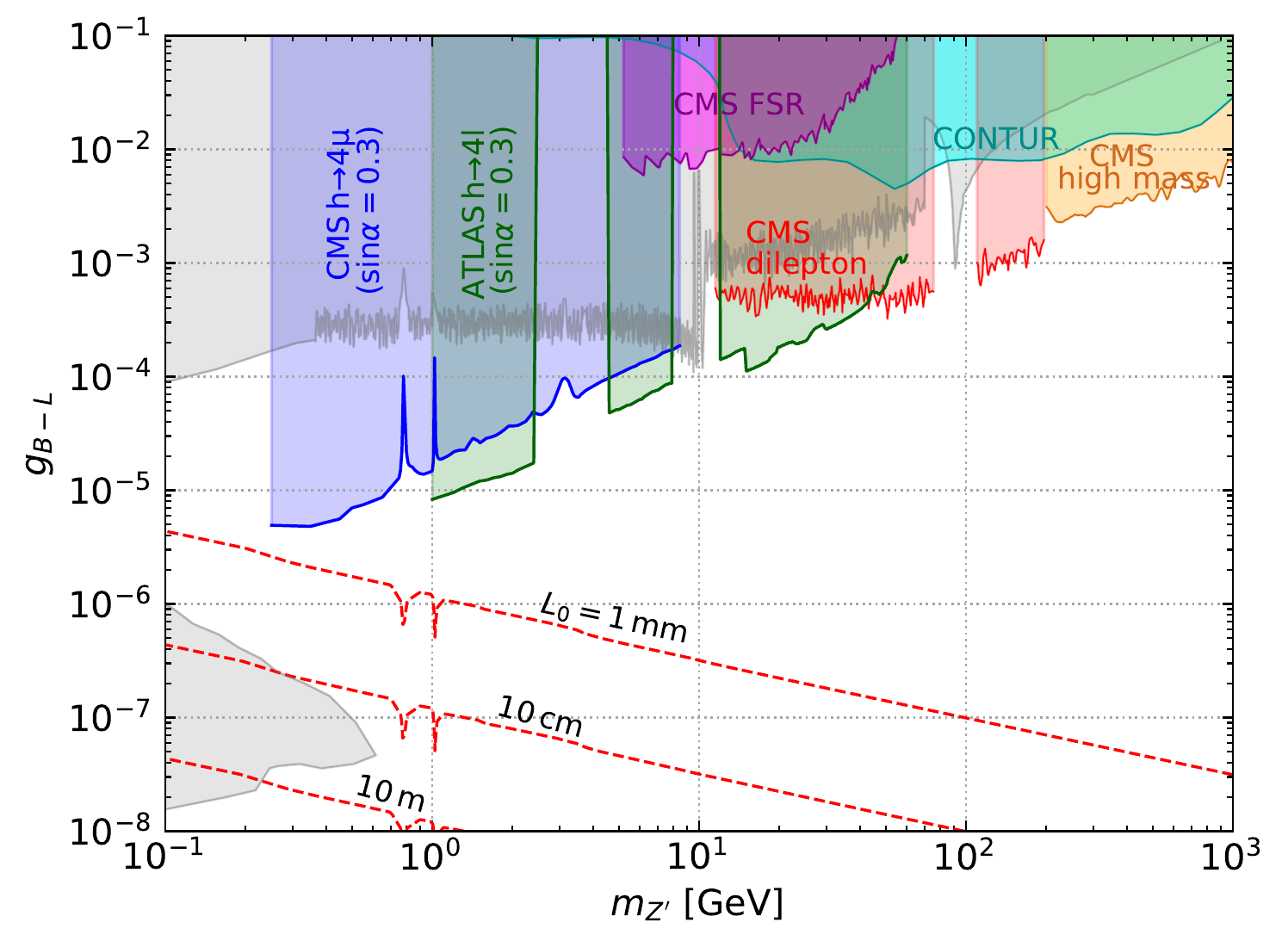}
	\caption{Summary of constraints on the $U(1)_{B-L}$ gauge coupling $g_{B-L}$ as a function of the $Z'$ mass $m_{Z'}$. The grey area represents existing constraints whereas the coloured regions represent the new constraints obtained in this paper derived from the CMS $h\to 4\mu$ \cite{Sirunyan:2018mgs}, ATLAS $h\to 4\ell $ \cite{Aaboud:2018fvk}, CMS FSR \cite{Sirunyan:2018nnz} and CMS dilepton \cite{CMS-PAS-EXO-19-018} searches. Also shown are the constraints derived from the CMS dilepton search \cite{CMS-PAS-EXO-19-019} (CMS high mass) and from LHC SM measurements using {\tt CONTUR} \cite{Amrith:2018yfb} (CONTUR). For the Higgs mediated modes we assume a Higgs mixing angle $\sin\alpha = 0.3$.}
	\label{fig:darkcast_summary}
\end{figure}
We have mainly explored three different \zprime production mechanisms at the LHC. The \zprime can either be produced via $s$-channel Drell-Yan, decays of the SM Higgs or via the final state radiation of the muons produced in the SM Drell-Yan process at the LHC. We demonstrated that the limits from the final state radiation arising from \cite{Sirunyan:2018nnz} are the weakest. As for the four-lepton final states produced via the SM Higgs we showed that the limits from existing searches are sensitive to the macroscopic \zprime displacements at the LHC. These searches have a potential to constrain large regions of \mzprime -- \gprime parameter space and are particularly powerful for light \zprime masses. The constraints from the ATLAS search~\cite{Aad:2019fac} are somewhat stronger than the CMS search~\cite{CMS-PAS-EXO-19-019} as ATLAS takes into account \zprime decays to both electrons and muon final state. On the other hand, the CMS search covers a wider $m_{Z'}$ interval. The Higgs production mode depends on the Higgs mixing angle $\sin\alpha$. In our analysis we have chosen a representative, approximately maximal value (given current limits) of $\sin\alpha = 0.3$. For smaller values the sensitivity to $g_{B-L}$ will accordingly weaken but our results illustrate the interplay of parameters in a realistic gauge model and the potential sensitivity to small exotic gauge couplings at the LHC. The four-lepton final state searches are further complemented by searches for dileptons. The most recent scouting analysis of the CMS dimuon search \cite{CMS-PAS-EXO-19-018} presents competitive limits in the mass range of 10 to 60~GeV.

The summary of our results is presented in Fig.~\ref{fig:darkcast_summary}. The plot also contains previously known limits on \zprime masses. In the mass region of 0.25 -- 1~GeV our analysis shows that there is a sensitivity improvement to probe $Z'$ limits by an order of magnitude. In the mass region between 10 to 60~GeV, the Higgs mediated channel and the recent CMS dilepton search in the dimuon final state also improve on the existing limits. Finally, for completeness, we also derive limits by interpreting recent high mass dilepton resonance searches. For this purpose, we use results from the most recent CMS high mass dilepton search \cite{CMS-PAS-EXO-19-019}. This analysis presents limits on the ratio of the dilepton resonance cross section to the SM $Z$ to dilepton production cross section. Taking the $Z$ to muon cross section to be 1870~pb, we derive limits on \gprime. We have checked that the corresponding ATLAS analysis \cite{Aad:2019fac}, yields similar limits. It should be noted that the limits from high mass resonance searches constrain masses well beyond 1~TeV.
The figure also illustrates the gaps in the dilepton resonance searches at ATLAS and CMS. In the region around \mzprime of 10~GeV, only weak constraints from the CMS FSR analysis can be derived, while the region around the $Z$ mass remains unconstrained by current LHC searches. As discussed in \cite{Amrith:2018yfb}, using the 'Constraints On New Theories Using Rivet' ({\tt CONTUR}) method of interpreting LHC SM measurements can still be used to extract constraints, albeit comparatively weaker. The resulting limits on $g_{B-L}$ are indicated in Fig.~\ref{fig:darkcast_summary}.

The ultimate prize when probing models such as the minimal $U(1)_{B-L}$ is to unravel the mechanism of neutrino mass generation. In our case this corresponds to discovering the heavy Majorana neutrinos giving rise to the seesaw mechanism. Because the heavy neutrinos are charged under the $U(1)_{B-L}$ gauge group, they can be produced not only via their mixing with the active neutrinos, which is generically expected to be small to explain the lightness of neutrinos, but also via the $Z'$, the $U(1)_{B-L}$ breaking Higgs $\chi$ and the SM Higgs (due to Higgs mixing). We here focus on the production of the $Z'$ at the LHC in the minimal $B-L$ models. Other aspects of $B-L$ models were discussed elsewhere. For example, a $B-L$ model with a specific low scale seesaw mechanism is discussed in Ref.~\cite{Khalil:2006yi}, with an inverse seesaw scenario in Ref.~\cite{Khalil:2010iu} and with a linear seesaw scenario in Ref.~\cite{Dib:2014fua}. Other aspects of heavy neutrinos were for example discussed in Refs.~\cite{Das:2017flq, Das:2017deo, Chun:2018ibr, Das:2018tbd, Jana:2018rdf}, including displaced vertex signatures. Finally, dark matter can be incorporated in $B-L$ models as well as has been for example discussed in Refs.~\cite{Klasen:2016qux, FileviezPerez:2019cyn, Heeba:2019jho, Mohapatra:2019ysk}.

The SM Higgs and $Z'$ channels were recently discussed in~\cite{Deppisch:2018eth, Deppisch:2019kvs, Das:2019fee, Chiang:2019ajm}. The vertex coupling the $hNN$ (SM Higgs) is proportional to $m_N/m_{Z^\prime}\times \sin\alpha\,g_{B-L}$, whereas the $Z'NN$ vertex is proportional to \gprime. From our analysis, we can thus infer new limits on the heavy neutrino production modes. As shown in Fig.~\ref{fig:limits_fixed_alpha}, when the \zprime production via Higgs is feasible, we obtain a conservative limit of $g_{B-L} < 10^{-4}$ for \zprime masses between 10 to 60~GeV. For lower masses of \zprime the limits get even more constraining. With this revised constraint on the \zprime coupling the heavy neutrino production via both channels is therefore suppressed and is not expected to yield a detectable cross section. Likewise, the branching ratio BR($h\to NN$) depends on \mzprime/\gprime; For $m_{Z'} < 100$~GeV and when applying our constraints this ratio is $\approx 100$~TeV. This is about 30 times larger than the value considered in Ref.~\cite{Deppisch:2018eth}, resulting in about a thousand times smaller cross section. Heavy neutrino production from $Z^\prime$ decays mentioned in Ref.~\cite{Deppisch:2019kvs} is suppressed as well due to roughly a magnitude better constraint on \gprime which makes the \zprime production cross section a hundred times smaller. This of course applies to the case where the Higgs mixing is near its maximally allowed value, $\sin\alpha\approx 0.3$, and the discussion will change for smaller values; in such a case the heavy neutrino Higgs portal $pp\to h\to NN$ will be suppressed though. With BR($Z^\prime\to NN$) $\approx 6\%$, the largest $pp\to Z^\prime \to NN$ cross section is only several femtobarn for a narrow range of \zprime masses between 10-15~GeV. One can also consider heavy neutrinos in cascade decays such as $pp\to h \to Z^\prime Z^\prime \to NN + X$. With $\sigma(pp\to h \to Z^\prime Z^\prime) \approx 1$~fb and BR($Z^\prime\to NN) \times \text{BR}(N\to \mu + X) \approx 1\%$ for $g_{B-L}\approx 10^{-5}$ and $m_{Z^\prime} > 1$~GeV, the total cross section of this process amounts to $10^{-2}$~fb. 

These considerations demonstrate that with the updated limits considered in this work, there is very little room for producing heavy neutrinos with long decay lengths (heavy neutrinos with shorter decay lengths can still be searched for via $W$ and $Z$ decays). It may still be possible to gain some sensitivity in this channel for the High Luminosity LHC. With these considerations we merely like to point out the importance of searching not only for heavy neutral leptons (i.e. heavy neutrinos) but also for the potential exotic mediators and portals through which they can be produced. This will shed light on whether the light neutrino masses have their origin in new physics around the TeV scale.

\acknowledgments

WL acknowledges support via the China Scholarship Council (Grant CSC No. 2016 08060325). SK is supported by Elise-Richter grant project number V592-N27 of the Austrian Science Fund and FFD by a UK STFC consolidated grant (Reference ST/P00072X/1). We thank Alberto Escalante del Valle and Ivan Mikulec (CMS) for several useful discussions.

\bibliographystyle{JHEP}
\bibliography{F-S-W-h1zpzp}

\providecommand{\href}[2]{#2}\begingroup\raggedright\begin{thebibliography}{10}

\bibitem{Mohapatra:1986bd}
R.~N. Mohapatra and J.~W.~F. Valle, \emph{{Neutrino Mass and Baryon Number
  Nonconservation in Superstring Models}},
  \href{http://dx.doi.org/10.1103/PhysRevD.34.1642}{\emph{Phys. Rev.} {\bf D34}
  (1986) 1642}.

\bibitem{Davidson:1978pm}
A.~Davidson, \emph{{$B^-$l as the Fourth Color, Quark - Lepton Correspondence,
  and Natural Masslessness of Neutrinos Within a Generalized Ws Model}},
  \href{http://dx.doi.org/10.1103/PhysRevD.20.776}{\emph{Phys. Rev.} {\bf D20}
  (1979) 776}.

\bibitem{Mohapatra:1980qe}
R.~N. Mohapatra and R.~E. Marshak, \emph{{Local B-L Symmetry of Electroweak
  Interactions, Majorana Neutrinos and Neutron Oscillations}},
  \href{http://dx.doi.org/10.1103/PhysRevLett.44.1644.2,
  10.1103/PhysRevLett.44.1316}{\emph{Phys. Rev. Lett.} {\bf 44} (1980)
  1316--1319}.

\bibitem{Aaboud:2017buh}
{\scshape ATLAS} collaboration, M.~Aaboud et~al., \emph{{Search for new
  high-mass phenomena in the dilepton final state using 36 fb$^{-1}$ of
  proton-proton collision data at $ \sqrt{s}=13 $ TeV with the ATLAS
  detector}}, \href{http://dx.doi.org/10.1007/JHEP10(2017)182}{\emph{JHEP} {\bf
  10} (2017) 182}, [\href{http://arxiv.org/abs/1707.02424}{{\tt 1707.02424}}].

\bibitem{LEP:2003aa}
{\scshape SLD Electroweak Group, SLD Heavy Flavor Group, DELPHI, LEP, ALEPH,
  OPAL, LEP Electroweak Working Group, L3} collaboration, t.~S. Electroweak,
  \emph{{A Combination of preliminary electroweak measurements and constraints
  on the standard model}},  \href{http://arxiv.org/abs/hep-ex/0312023}{{\tt
  hep-ex/0312023}}.

\bibitem{Anthony:2003ub}
{\scshape SLAC E158} collaboration, P.~L. Anthony et~al., \emph{{Observation of
  parity nonconservation in Moller scattering}},
  \href{http://dx.doi.org/10.1103/PhysRevLett.92.181602}{\emph{Phys. Rev.
  Lett.} {\bf 92} (2004) 181602},
  [\href{http://arxiv.org/abs/hep-ex/0312035}{{\tt hep-ex/0312035}}].

\bibitem{Carena:2004xs}
M.~Carena, A.~Daleo, B.~A. Dobrescu and T.~M.~P. Tait, \emph{{$Z^\prime$ gauge
  bosons at the Tevatron}},
  \href{http://dx.doi.org/10.1103/PhysRevD.70.093009}{\emph{Phys. Rev.} {\bf
  D70} (2004) 093009}, [\href{http://arxiv.org/abs/hep-ph/0408098}{{\tt
  hep-ph/0408098}}].

\bibitem{Cacciapaglia:2006pk}
G.~Cacciapaglia, C.~Csaki, G.~Marandella and A.~Strumia, \emph{{The Minimal Set
  of Electroweak Precision Parameters}},
  \href{http://dx.doi.org/10.1103/PhysRevD.74.033011}{\emph{Phys. Rev.} {\bf
  D74} (2006) 033011}, [\href{http://arxiv.org/abs/hep-ph/0604111}{{\tt
  hep-ph/0604111}}].

\bibitem{Harnik:2012ni}
R.~Harnik, J.~Kopp and P.~A.~N. Machado, \emph{{Exploring nu Signals in Dark
  Matter Detectors}},
  \href{http://dx.doi.org/10.1088/1475-7516/2012/07/026}{\emph{JCAP} {\bf 1207}
  (2012) 026}, [\href{http://arxiv.org/abs/1202.6073}{{\tt 1202.6073}}].

\bibitem{Bellini:2011rx}
G.~Bellini et~al., \emph{{Precision measurement of the 7Be solar neutrino
  interaction rate in Borexino}},
  \href{http://dx.doi.org/10.1103/PhysRevLett.107.141302}{\emph{Phys. Rev.
  Lett.} {\bf 107} (2011) 141302}, [\href{http://arxiv.org/abs/1104.1816}{{\tt
  1104.1816}}].

\bibitem{Bauer:2018onh}
M.~Bauer, P.~Foldenauer and J.~Jaeckel, \emph{{Hunting All the Hidden
  Photons}}, \href{http://dx.doi.org/10.1007/JHEP07(2018)094}{\emph{JHEP} {\bf
  07} (2018) 094}, [\href{http://arxiv.org/abs/1803.05466}{{\tt 1803.05466}}].

\bibitem{Vilain:1993kd}
{\scshape CHARM-II} collaboration, P.~Vilain et~al., \emph{{Measurement of
  differential cross-sections for muon-neutrino electron scattering}},
  \href{http://dx.doi.org/10.1016/0370-2693(93)90408-A}{\emph{Phys. Lett.} {\bf
  B302} (1993) 351--355}.

\bibitem{Deniz:2009mu}
{\scshape TEXONO} collaboration, M.~Deniz et~al., \emph{{Measurement of
  Nu(e)-bar -Electron Scattering Cross-Section with a CsI(Tl) Scintillating
  Crystal Array at the Kuo-Sheng Nuclear Power Reactor}},
  \href{http://dx.doi.org/10.1103/PhysRevD.81.072001}{\emph{Phys. Rev.} {\bf
  D81} (2010) 072001}, [\href{http://arxiv.org/abs/0911.1597}{{\tt
  0911.1597}}].

\bibitem{Amrith:2018yfb}
S.~Amrith, J.~M. Butterworth, F.~F. Deppisch, W.~Liu, A.~Varma and D.~Yallup,
  \emph{{LHC Constraints on a $B-L$ Gauge Model using Contur}},
  \href{http://arxiv.org/abs/1811.11452}{{\tt 1811.11452}}.

\bibitem{Butterworth:2016sqg}
J.~M. Butterworth, D.~Grellscheid, M.~Kr$\ddot{\text{a}}$mer, B.~Sarrazin and
  D.~Yallup, \emph{{Constraining new physics with collider measurements of
  Standard Model signatures}},
  \href{http://dx.doi.org/10.1007/JHEP03(2017)078}{\emph{JHEP} {\bf 03} (2017)
  078}, [\href{http://arxiv.org/abs/1606.05296}{{\tt 1606.05296}}].

\bibitem{Ilten:2018crw}
P.~Ilten, Y.~Soreq, M.~Williams and W.~Xue, \emph{{Serendipity in dark photon
  searches}}, \href{http://dx.doi.org/10.1007/JHEP06(2018)004}{\emph{JHEP} {\bf
  06} (2018) 004}, [\href{http://arxiv.org/abs/1801.04847}{{\tt 1801.04847}}].

\bibitem{Aaij:2017rft}
{\scshape LHCb} collaboration, R.~Aaij et~al., \emph{{Search for Dark Photons
  Produced in 13 TeV $pp$ Collisions}},
  \href{http://dx.doi.org/10.1103/PhysRevLett.120.061801}{\emph{Phys. Rev.
  Lett.} {\bf 120} (2018) 061801}, [\href{http://arxiv.org/abs/1710.02867}{{\tt
  1710.02867}}].

\bibitem{Bechtle:2013xfa}
P.~Bechtle, S.~Heinemeyer, O.~Stal, T.~Stefaniak and G.~Weiglein,
  \emph{{$HiggsSignals$: Confronting arbitrary Higgs sectors with measurements
  at the Tevatron and the LHC}},
  \href{http://dx.doi.org/10.1140/epjc/s10052-013-2711-4}{\emph{Eur. Phys. J.}
  {\bf C74} (2014) 2711}, [\href{http://arxiv.org/abs/1305.1933}{{\tt
  1305.1933}}].

\bibitem{Ilnicka:2018def}
A.~Ilnicka, T.~Robens and T.~Stefaniak, \emph{{Constraining Extended Scalar
  Sectors at the LHC and beyond}},
  \href{http://dx.doi.org/10.1142/S0217732318300070}{\emph{Mod. Phys. Lett.}
  {\bf A33} (2018) 1830007}, [\href{http://arxiv.org/abs/1803.03594}{{\tt
  1803.03594}}].

\bibitem{Bechtle:2014ewa}
P.~Bechtle, S.~Heinemeyer, O.~Stal, T.~Stefaniak and G.~Weiglein,
  \emph{{Probing the Standard Model with Higgs signal rates from the Tevatron,
  the LHC and a future ILC}},
  \href{http://dx.doi.org/10.1007/JHEP11(2014)039}{\emph{JHEP} {\bf 11} (2014)
  039}, [\href{http://arxiv.org/abs/1403.1582}{{\tt 1403.1582}}].

\bibitem{ATLAS:2016gld}
{\scshape ATLAS} collaboration, \emph{{Measurements of the Higgs boson
  production cross section via Vector Boson Fusion and associated $WH$
  production in the $WW^{\ast} \to \ell\nu\ell\nu$ decay mode with the ATLAS
  detector at $\sqrt{s}$ = 13 TeV}}, .

\bibitem{ATLAS:2016oum}
{\scshape ATLAS} collaboration, \emph{{Study of the Higgs boson properties and
  search for high-mass scalar resonances in the $H \rightarrow ZZ^* \rightarrow
  4\ell$ decay channel at $\sqrt{s}$ = 13 TeV with the ATLAS detector}}, .

\bibitem{ATLAS:2016pkl}
{\scshape ATLAS} collaboration, \emph{{Search for the Standard Model Higgs
  boson produced in association with a vector boson and decaying to a
  $b\bar{b}$ pair in $pp$ collisions at 13 TeV using the ATLAS detector}}, .

\bibitem{ATLAS:2016awy}
{\scshape ATLAS} collaboration, \emph{{Search for the Standard Model Higgs
  boson produced in association with top quarks and decaying into
  $b\overline{b}$ in $pp$ collisions at $\sqrt{s}$ = 13 TeV with the ATLAS
  detector}}, .

\bibitem{ATLAS:2016nke}
{\scshape ATLAS} collaboration, \emph{{Measurement of fiducial, differential
  and production cross sections in the $H\to\gamma\gamma$ decay channel with
  13.3 fb$^{-1}$ of 13 TeV proton-proton collision data with the ATLAS
  detector}}, .

\bibitem{ATLAS:2016ldo}
{\scshape ATLAS} collaboration, \emph{{Search for the Associated Production of
  a Higgs Boson and a Top Quark Pair in Multilepton Final States with the ATLAS
  Detector}}, .

\bibitem{Sirunyan:2017exp}
{\scshape CMS} collaboration, A.~M. Sirunyan et~al., \emph{{Measurements of
  properties of the Higgs boson decaying into the four-lepton final state in pp
  collisions at $ \sqrt{s}=13 $ TeV}},
  \href{http://dx.doi.org/10.1007/JHEP11(2017)047}{\emph{JHEP} {\bf 11} (2017)
  047}, [\href{http://arxiv.org/abs/1706.09936}{{\tt 1706.09936}}].

\bibitem{CMS:2016ixj}
{\scshape CMS} collaboration, \emph{{Updated measurements of Higgs boson
  production in the diphoton decay channel at $\sqrt{s}=13~\textrm{TeV}$ in pp
  collisions at CMS.}}, .

\bibitem{CMS:aya}
{\scshape CMS} collaboration, \emph{{Combination of standard model Higgs boson
  searches and measurements of the properties of the new boson with a mass near
  125 GeV}}, .

\bibitem{CMS:bxa}
{\scshape CMS} collaboration, \emph{{Update on the search for the standard
  model Higgs boson in pp collisions at the LHC decaying to W + W in the fully
  leptonic final state}}, .

\bibitem{Khachatryan:2015cwa}
{\scshape CMS} collaboration, V.~Khachatryan et~al., \emph{{Search for a Higgs
  boson in the mass range from 145 to 1000 GeV decaying to a pair of W or Z
  bosons}}, \href{http://dx.doi.org/10.1007/JHEP10(2015)144}{\emph{JHEP} {\bf
  10} (2015) 144}, [\href{http://arxiv.org/abs/1504.00936}{{\tt 1504.00936}}].

\bibitem{CMS:2017vpy}
{\scshape CMS} collaboration, C.~Collaboration, \emph{{Search for a new scalar
  resonance decaying to a pair of Z bosons in proton-proton collisions at
  $\sqrt s$ = 13 TeV}}, .

\bibitem{Aaboud:2017rel}
{\scshape ATLAS} collaboration, M.~Aaboud et~al., \emph{{Search for heavy ZZ
  resonances in the $\ell ^+\ell ^-\ell ^+\ell ^-$ and $\ell ^+\ell ^-\nu
  \bar{\nu }$ final states using proton–proton collisions at $\sqrt{s}= 13$
  $\text {TeV}$ with the ATLAS detector}},
  \href{http://dx.doi.org/10.1140/epjc/s10052-018-5686-3}{\emph{Eur. Phys. J.}
  {\bf C78} (2018) 293}, [\href{http://arxiv.org/abs/1712.06386}{{\tt
  1712.06386}}].

\bibitem{Lopez-Val:2014jva}
D.~Lopez-Val and T.~Robens, \emph{{$\Delta$r and the W-boson mass in the
  singlet extension of the standard model}},
  \href{http://dx.doi.org/10.1103/PhysRevD.90.114018}{\emph{Phys. Rev.} {\bf
  D90} (2014) 114018}, [\href{http://arxiv.org/abs/1406.1043}{{\tt
  1406.1043}}].

\bibitem{Robens:2015gla}
T.~Robens and T.~Stefaniak, \emph{{Status of the Higgs Singlet Extension of the
  Standard Model after LHC Run 1}},
  \href{http://dx.doi.org/10.1140/epjc/s10052-015-3323-y}{\emph{Eur. Phys. J.}
  {\bf C75} (2015) 104}, [\href{http://arxiv.org/abs/1501.02234}{{\tt
  1501.02234}}].

\bibitem{DEramo:2017zqw}
F.~D'Eramo, B.~J. Kavanagh and P.~Panci, \emph{{Probing Leptophilic Dark
  Sectors with Hadronic Processes}},
  \href{http://dx.doi.org/10.1016/j.physletb.2017.05.063}{\emph{Phys. Lett.}
  {\bf B771} (2017) 339--348}, [\href{http://arxiv.org/abs/1702.00016}{{\tt
  1702.00016}}].

\bibitem{Anastasi:2015qla}
A.~Anastasi et~al., \emph{{Limit on the production of a low-mass vector boson
  in $\mathrm{e}^{+}\mathrm{e}^{-} \to \mathrm{U}\gamma$, $\mathrm{U} \to
  \mathrm{e}^{+}\mathrm{e}^{-}$ with the KLOE experiment}},
  \href{http://dx.doi.org/10.1016/j.physletb.2015.10.003}{\emph{Phys. Lett.}
  {\bf B750} (2015) 633--637}, [\href{http://arxiv.org/abs/1509.00740}{{\tt
  1509.00740}}].

\bibitem{Lees:2014xha}
{\scshape BaBar} collaboration, J.~P. Lees et~al., \emph{{Search for a Dark
  Photon in $e^+e^-$ Collisions at BaBar}},
  \href{http://dx.doi.org/10.1103/PhysRevLett.113.201801}{\emph{Phys. Rev.
  Lett.} {\bf 113} (2014) 201801}, [\href{http://arxiv.org/abs/1406.2980}{{\tt
  1406.2980}}].

\bibitem{Sirunyan:2018exx}
{\scshape CMS} collaboration, A.~M. Sirunyan et~al., \emph{{Search for
  high-mass resonances in dilepton final states in proton-proton collisions at
  $\sqrt{s}=$ 13 TeV}},
  \href{http://dx.doi.org/10.1007/JHEP06(2018)120}{\emph{JHEP} {\bf 06} (2018)
  120}, [\href{http://arxiv.org/abs/1803.06292}{{\tt 1803.06292}}].

\bibitem{Aad:2019fac}
{\scshape ATLAS} collaboration, G.~Aad et~al., \emph{{Search for high-mass
  dilepton resonances using 139 fb$^{-1}$ of $pp$ collision data collected at
  $\sqrt{s}=$13 TeV with the ATLAS detector}},
  \href{http://dx.doi.org/10.1016/j.physletb.2019.07.016}{\emph{Phys. Lett.}
  {\bf B796} (2019) 68--87}, [\href{http://arxiv.org/abs/1903.06248}{{\tt
  1903.06248}}].

\bibitem{Sirunyan:2019sgo}
{\scshape CMS} collaboration, A.~M. Sirunyan et~al., \emph{{Search for low-mass
  quark-antiquark resonances produced in association with a photon at $\sqrt{s}
  = $ 13 TeV}},  \href{http://arxiv.org/abs/1905.10331}{{\tt 1905.10331}}.

\bibitem{Gu:2017ckc}
J.~Gu, H.~Li, Z.~Liu, S.~Su and W.~Su, \emph{{Learning from Higgs Physics at
  Future Higgs Factories}},
  \href{http://dx.doi.org/10.1007/JHEP12(2017)153}{\emph{JHEP} {\bf 12} (2017)
  153}, [\href{http://arxiv.org/abs/1709.06103}{{\tt 1709.06103}}].

\bibitem{CEPCStudyGroup:2018ghi}
{\scshape CEPC Study Group} collaboration, M.~Dong et~al., \emph{{CEPC
  Conceptual Design Report: Volume 2 - Physics \& Detector}},
  \href{http://arxiv.org/abs/1811.10545}{{\tt 1811.10545}}.

\bibitem{Fujiwara:1984mp}
T.~Fujiwara, T.~Kugo, H.~Terao, S.~Uehara and K.~Yamawaki, \emph{{Nonabelian
  Anomaly and Vector Mesons as Dynamical Gauge Bosons of Hidden Local
  Symmetries}}, \href{http://dx.doi.org/10.1143/PTP.73.926}{\emph{Prog. Theor.
  Phys.} {\bf 73} (1985) 926}.

\bibitem{CMS-PAS-EXO-19-018}
{\scshape CMS Collaboration} collaboration, C.~collaboration, \emph{{Search for
  a narrow resonance decaying to a pair of muons in proton-proton collisions at
  13 TeV}},  Tech. Rep. CMS-PAS-EXO-19-018, CERN, Geneva, 2019.

\bibitem{Sirunyan:2018mgs}
{\scshape CMS} collaboration, A.~M. Sirunyan et~al., \emph{{A search for pair
  production of new light bosons decaying into muons in proton-proton
  collisions at 13 TeV}}, {\emph{Submitted to: Phys. Lett.} (2018) },
  [\href{http://arxiv.org/abs/1812.00380}{{\tt 1812.00380}}].

\bibitem{Aaboud:2018fvk}
{\scshape ATLAS} collaboration, M.~Aaboud et~al., \emph{{Search for Higgs boson
  decays to beyond-the-Standard-Model light bosons in four-lepton events with
  the ATLAS detector at $\sqrt{s}=13$ TeV}},
  \href{http://dx.doi.org/10.1007/JHEP06(2018)166}{\emph{JHEP} {\bf 06} (2018)
  166}, [\href{http://arxiv.org/abs/1802.03388}{{\tt 1802.03388}}].

\bibitem{Sirunyan:2018nnz}
{\scshape CMS} collaboration, A.~M. Sirunyan et~al., \emph{{Search for an
  $L_{\mu}-L_{\tau}$ gauge boson using Z$\to4\mu$ events in proton-proton
  collisions at $\sqrt{s} =$ 13 TeV}},
  \href{http://dx.doi.org/10.1016/j.physletb.2019.01.072}{\emph{Phys. Lett.}
  {\bf B792} (2019) 345--368}, [\href{http://arxiv.org/abs/1808.03684}{{\tt
  1808.03684}}].

\bibitem{deFlorian:2016spz}
{\scshape LHC Higgs Cross Section Working Group} collaboration, D.~de~Florian
  et~al., \emph{{Handbook of LHC Higgs Cross Sections: 4. Deciphering the
  Nature of the Higgs Sector}},  \href{http://arxiv.org/abs/1610.07922}{{\tt
  1610.07922}}.

\bibitem{Tanabashi:2018oca}
{\scshape Particle Data Group} collaboration, M.~Tanabashi et~al.,
  \emph{{Review of Particle Physics}},
  \href{http://dx.doi.org/10.1103/PhysRevD.98.030001}{\emph{Phys. Rev.} {\bf
  D98} (2018) 030001}.

\bibitem{Deppisch:2018eth}
F.~F. Deppisch, W.~Liu and M.~Mitra, \emph{{Long-lived Heavy Neutrinos from
  Higgs Decays}}, \href{http://dx.doi.org/10.1007/JHEP08(2018)181}{\emph{JHEP}
  {\bf 08} (2018) 181}, [\href{http://arxiv.org/abs/1804.04075}{{\tt
  1804.04075}}].

\bibitem{Degrande:2011ua}
C.~Degrande, C.~Duhr, B.~Fuks, D.~Grellscheid, O.~Mattelaer and T.~Reiter,
  \emph{{UFO - The Universal FeynRules Output}},
  \href{http://dx.doi.org/10.1016/j.cpc.2012.01.022}{\emph{Comput. Phys.
  Commun.} {\bf 183} (2012) 1201--1214},
  [\href{http://arxiv.org/abs/1108.2040}{{\tt 1108.2040}}].

\bibitem{Alwall:2014hca}
J.~Alwall, R.~Frederix, S.~Frixione, V.~Hirschi, F.~Maltoni, O.~Mattelaer
  et~al., \emph{{The automated computation of tree-level and next-to-leading
  order differential cross sections, and their matching to parton shower
  simulations}}, \href{http://dx.doi.org/10.1007/JHEP07(2014)079}{\emph{JHEP}
  {\bf 07} (2014) 079}, [\href{http://arxiv.org/abs/1405.0301}{{\tt
  1405.0301}}].

\bibitem{Sjostrand:2014zea}
T.~Sj$\ddot{\text{o}}$strand, S.~Ask, J.~R. Christiansen, R.~Corke, N.~Desai,
  P.~Ilten et~al., \emph{{An Introduction to PYTHIA 8.2}},
  \href{http://dx.doi.org/10.1016/j.cpc.2015.01.024}{\emph{Comput. Phys.
  Commun.} {\bf 191} (2015) 159--177},
  [\href{http://arxiv.org/abs/1410.3012}{{\tt 1410.3012}}].

\bibitem{Aad:2015sms}
{\scshape ATLAS} collaboration, G.~Aad et~al., \emph{{A search for prompt
  lepton-jets in $pp$ collisions at $\sqrt{s}=$ 8 TeV with the ATLAS
  detector}}, \href{http://dx.doi.org/10.1007/JHEP02(2016)062}{\emph{JHEP} {\bf
  02} (2016) 062}, [\href{http://arxiv.org/abs/1511.05542}{{\tt 1511.05542}}].

\bibitem{Aad:2014yea}
{\scshape ATLAS} collaboration, G.~Aad et~al., \emph{{Search for long-lived
  neutral particles decaying into lepton jets in proton-proton collisions at $
  \sqrt{s}=8 $ TeV with the ATLAS detector}},
  \href{http://dx.doi.org/10.1007/JHEP11(2014)088}{\emph{JHEP} {\bf 11} (2014)
  088}, [\href{http://arxiv.org/abs/1409.0746}{{\tt 1409.0746}}].

\bibitem{Aad:2019tua}
{\scshape ATLAS} collaboration, G.~Aad et~al., \emph{{Search for light
  long-lived neutral particles produced in $pp$ collisions at $\sqrt{s} =$ 13
  TeV and decaying into collimated leptons or light hadrons with the ATLAS
  detector}},  \href{http://arxiv.org/abs/1909.01246}{{\tt 1909.01246}}.

\bibitem{CMS-PAS-EXO-19-019}
{\scshape CMS Collaboration} collaboration, \emph{{Search for a narrow
  resonance in high-mass dilepton final states in proton-proton collisions
  using 140$~\mathrm{fb}^{-1}$ of data at $\sqrt{s}=13~\mathrm{TeV}$}},  Tech.
  Rep. CMS-PAS-EXO-19-019, CERN, Geneva, 2019.

\bibitem{Khalil:2006yi}
S.~Khalil, \emph{{Low scale $B$ - L extension of the Standard Model at the
  LHC}}, \href{http://dx.doi.org/10.1088/0954-3899/35/5/055001}{\emph{J. Phys.}
  {\bf G35} (2008) 055001}, [\href{http://arxiv.org/abs/hep-ph/0611205}{{\tt
  hep-ph/0611205}}].

\bibitem{Khalil:2010iu}
S.~Khalil, \emph{{TeV-scale gauged B-L symmetry with inverse seesaw
  mechanism}}, \href{http://dx.doi.org/10.1103/PhysRevD.82.077702}{\emph{Phys.
  Rev.} {\bf D82} (2010) 077702}, [\href{http://arxiv.org/abs/1004.0013}{{\tt
  1004.0013}}].

\bibitem{Dib:2014fua}
C.~O. Dib, G.~R. Moreno and N.~A. Neill, \emph{{Neutrinos with a linear seesaw
  mechanism in a scenario of gauged B-L symmetry}},
  \href{http://dx.doi.org/10.1103/PhysRevD.90.113003}{\emph{Phys. Rev.} {\bf
  D90} (2014) 113003}, [\href{http://arxiv.org/abs/1409.1868}{{\tt
  1409.1868}}].

\bibitem{Das:2017flq}
A.~Das, N.~Okada and D.~Raut, \emph{{Enhanced pair production of heavy Majorana
  neutrinos at the LHC}},
  \href{http://dx.doi.org/10.1103/PhysRevD.97.115023}{\emph{Phys. Rev.} {\bf
  D97} (2018) 115023}, [\href{http://arxiv.org/abs/1710.03377}{{\tt
  1710.03377}}].

\bibitem{Das:2017deo}
A.~Das, N.~Okada and D.~Raut, \emph{{Heavy Majorana neutrino pair productions
  at the LHC in minimal U(1) extended Standard Model}},
  \href{http://dx.doi.org/10.1140/epjc/s10052-018-6171-8}{\emph{Eur. Phys. J.}
  {\bf C78} (2018) 696}, [\href{http://arxiv.org/abs/1711.09896}{{\tt
  1711.09896}}].

\bibitem{Chun:2018ibr}
E.~J. Chun, A.~Das, J.~Kim and J.~Kim, \emph{{Searching for flavored gauge
  bosons}}, \href{http://dx.doi.org/10.1007/JHEP07(2019)024,
  10.1007/JHEP02(2019)093}{\emph{JHEP} {\bf 02} (2019) 093},
  [\href{http://arxiv.org/abs/1811.04320}{{\tt 1811.04320}}].

\bibitem{Das:2018tbd}
A.~Das, N.~Okada, S.~Okada and D.~Raut, \emph{{Probing the seesaw mechanism at
  the 250 GeV ILC}},  \href{http://arxiv.org/abs/1812.11931}{{\tt 1812.11931}}.

\bibitem{Jana:2018rdf}
S.~Jana, N.~Okada and D.~Raut, \emph{{Displaced vertex signature of type-I
  seesaw model}},
  \href{http://dx.doi.org/10.1103/PhysRevD.98.035023}{\emph{Phys. Rev.} {\bf
  D98} (2018) 035023}, [\href{http://arxiv.org/abs/1804.06828}{{\tt
  1804.06828}}].

\bibitem{Klasen:2016qux}
M.~Klasen, F.~Lyonnet and F.~S. Queiroz, \emph{{NLO+NLL collider bounds, Dirac
  fermion and scalar dark matter in the B–L model}},
  \href{http://dx.doi.org/10.1140/epjc/s10052-017-4904-8}{\emph{Eur. Phys. J.}
  {\bf C77} (2017) 348}, [\href{http://arxiv.org/abs/1607.06468}{{\tt
  1607.06468}}].

\bibitem{FileviezPerez:2019cyn}
P.~Fileviez~Pérez, C.~Murgui and A.~D. Plascencia, \emph{{Neutrino-Dark Matter
  Connections in Gauge Theories}},
  \href{http://dx.doi.org/10.1103/PhysRevD.100.035041}{\emph{Phys. Rev.} {\bf
  D100} (2019) 035041}, [\href{http://arxiv.org/abs/1905.06344}{{\tt
  1905.06344}}].

\bibitem{Heeba:2019jho}
S.~Heeba and F.~Kahlhoefer, \emph{{Probing the freeze-in mechanism in dark
  matter models with $U(1)^\prime$ gauge extensions}},
  \href{http://arxiv.org/abs/1908.09834}{{\tt 1908.09834}}.

\bibitem{Mohapatra:2019ysk}
R.~N. Mohapatra and N.~Okada, \emph{{Dark Matter Constraints on Low Mass and
  Weakly Coupled B-L Gauge Boson}},
  \href{http://arxiv.org/abs/1908.11325}{{\tt 1908.11325}}.

\bibitem{Deppisch:2019kvs}
F.~Deppisch, S.~Kulkarni and W.~Liu, \emph{{Heavy neutrino production via $Z'$
  at the lifetime frontier}},
  \href{http://dx.doi.org/10.1103/PhysRevD.100.035005}{\emph{Phys. Rev.} {\bf
  D100} (2019) 035005}, [\href{http://arxiv.org/abs/1905.11889}{{\tt
  1905.11889}}].

\bibitem{Das:2019fee}
A.~Das, P.~S.~B. Dev and N.~Okada, \emph{{Long-Lived TeV-Scale Right-Handed
  Neutrino Production at the LHC in Gauged $U(1)_X$ Model}},
  \href{http://arxiv.org/abs/1906.04132}{{\tt 1906.04132}}.

\bibitem{Chiang:2019ajm}
C.-W. Chiang, G.~Cottin, A.~Das and S.~Mandal, \emph{{Displaced heavy neutrinos
  from $Z'$ decays at the LHC}},  \href{http://arxiv.org/abs/1908.09838}{{\tt
  1908.09838}}.

\end{thebibliography}\endgroup
\end{document}